\documentclass[preprint,12pt]{elsarticle}
\usepackage{graphicx,psfrag,color}% Include figure files
\usepackage{mathbbol,amsmath,amsfonts,amssymb,bm,dsfont}
\usepackage{wasysym}
\usepackage{amsthm}

\def\dual{\ \stackrel{\Phi_\d}{\longrightarrow}\ }

\def\im{{\sf {i}}}
\def\d{{{\sf d}}}
\def\r{{\bm{r}}}
\def\x{{\bm{x}}}
\def\i{{\bm{e_1}}}
\def\j{{\bm{e_2}}}

\def\Z{{\mathbb{Z}}}
\def\one{{\mathds{1}}}
\def\tr{{{\sf Tr}\ }}

\def\pf{\mathcal{Z}}
\def\Z{{\mathbb{Z}}}
\def\tr{{\rm tr}\; }

\def\one{{\mathchoice {\rm 1\mskip-4mu l} {\rm 1\mskip-4mu l} {\rm
1\mskip-4.5mu l} {\rm 1\mskip-5mu l}}}

\newcommand{\half}{\mbox{$\textstyle \frac{1}{2}$} }
\newcommand{\ket}[1]{| #1 \rangle}

\journal{Nuclear Physics B}

\makeindex
\begin{document}
\begin{frontmatter}

\title{Dualities and the phase diagram of the $p$-clock model}

\author[]{G. Ortiz$^{1,}$\footnote{Corresponding author: ortizg@indiana.edu}, 
E. Cobanera$^1$, and Z. Nussinov$^2$}
%\index[aindx]{Author, F.} % or \aindx{Author, F.}
%\index[aindx]{Author, S.} % or \aindx{Author, S.}

\address{$^1$Department of Physics, Indiana University, Bloomington,
IN 47405, USA, \\ 
$^2$Department of Physics, Washington University, St.
Louis, MO 63160, USA.}

\begin{abstract}
A new ``bond-algebraic'' approach to duality transformations provides a
very powerful technique to analyze elementary excitations in the
classical two-dimensional XY and $p$-clock models. By combining duality
and Peierls arguments, we establish the existence of  non-Abelian
symmetries, the phase structure, and transitions of these models, unveil
the  nature of their topological excitations, and explicitly show that a
continuous U(1) symmetry emerges when $p \geq 5$. This latter symmetry
is associated with the appearance of discrete vortices and
Berezinskii-Kosterlitz-Thouless-type transitions. We derive a correlation 
inequality to prove that the intermediate phase, appearing for $p\geq
5$, is  critical (massless) with decaying power-law correlations.
\end{abstract}

\end{frontmatter}

\section{Introduction}

In this article we investigate, via the use of {\it dualities}, 
two-dimensional  ($D=2$) classical  systems, such as the  XY and clock
models  \cite{Potts, wu}, that display   Berezinskii-Kosterlitz-Thouless
(BKT)-type transitions \cite{berez,KT,K,Jose}.  BKT transitions, notably
characterized by essential singularities in  the free energy,  emerge in
many physical situations including screening in Coulomb gases
\cite{cg},    surface roughening \cite{sos},  melting in $D=2$ solids
\cite{KT,2dsolid},   and many other classical and quantum  problems,
such as deconfinement   in $D=3+1$ lattice gauge theories 
\cite{Elitzur,NO,ours1,dualities}. We study  these models by invoking a
{\it bond-algebraic}   approach that we have recently developed
\cite{ours1,dualities,ours2,ours0}. Within our duality-based method, one
relates singularities in the free energy at one temperature (or coupling
constants) to those at a dual temperature (or dual coupling constants). 

Specifically, we investigate exact dualities of the $D=2$ XY  and
$p$-clock models, and exploit those dualities to unravel their phase 
structures.  Those transformations are exact even for finite systems 
after appropriate boundary terms are included. It is noteworthy that
unlike nearly all of the analytical work done to  date, our dualities do
not rely on the   approximation scheme of Villain \cite{villain,
Elitzur, Savit, Einhorn}, yet they can be related to the exact dualities
of the Villain model in appropriate  limits.  Furthermore, our analysis leads
to exact  dualities for general $p$-clock models and yields a better
understanding of the appearance of two transitions in systems with $p
\ge 5$ states (the XY model with only one transition  is recovered in
the  $p \rightarrow \infty$ limit).   
%Amongst the many 
%rigorous results to be provided here, insofar as the phase diagram of
%these systems is concerned 
By fusing our duality  results with the {\it Peierls argument}, we will
be able to (1) prove that $p \ge 5$ clock systems can be made to be
self-dual; (2) prove by a Peierls argument that there exists a lower
ordering temperature $T^{(1)} \sim 1/p^{2}$, associated to domain-wall
excitations, below which the global $\mathbb{Z}_p$ symmetry is broken;
(3) demonstrate that  a second transition occurs at a temperature
$T^{(2)} \sim {\cal{O}}(1)$ when $p \geq 5$ (in the  self-dual case, it
follows that if $T^{(1)}$ is not the self-dual  temperature $T^*$, then
there must necessarily be a second phase transition  at $T^{(2)}$ of an
identical character); and (4) characterize the nature of the topological
excitations, and further explain that at $p=5$ a new type of topological
excitation, with an associated discrete  winding number, appears.   Our
considerations suggest that these {\it discrete vortices} may
proliferate above the temperature $T^{(2)}$. We will also (5) determine
an analytic expression for the self-dual temperature $T^*$,  which
relates and clarifies temperature scales discussed in \cite{Lapilli},
and most importantly, (6) establish the  non-Abelian polyhedral symmetry
group $P(2,2,p)$ of the $p$-clock  and related models, and explicitly
unveil the U(1) continuous symmetry that  {\it emerges} when $p \geq 5$.
Indeed, the latter is intimately tied to  the existence of the BKT
transition. Finally, (7) we derive a correlation  inequality to prove
that the intermediate phase, appearing for $p\geq 5$, is  critical
(massless) with decaying power-law correlations.

Despite several analytic \cite{Elitzur, Cardy, frohlich} and numerical
calculations   \cite{Borisenko,Lapilli,tomita}, the precise nature of
the two phase transitions  ($p \geq 5$) is not completely understood. It
was proven \cite{frohlich}  that for large enough $p$, clock  models
exhibit a BKT-type transition (actually, it has only been proved that 
there exists an intermediate critical phase with power-law
correlations). The question whether that still holds when $p=5$ remains
open  \cite{Borisenko}. By relying on exact results, we shed light on
the character  of the BKT transitions in these systems.  We will relate
the BKT transition to a continuous U(1) emergent symmetry of an usual
type.  Although BKT transitions are  often discussed in terms of
specific anomalous exponents and jumps in the helicity modulus, we will
not address such non-universal issues.  

Our treatment of classical dualities is based on a new approach
developed in Refs. \cite{ours2,dualities} that relies on the  transfer
matrix or operator formalism \cite{NO}. In statistical mechanics,  two
models $a$ and $b$  are dual if their partition functions \(\pf_a=\tr
[T_a^N]\) and \(\pf_b=\tr [T_b^N]\) are related as  ($N$ is the linear
size of the system in $D=2$)
\begin{equation}\label{class_d}
\pf_a[K]=A(K,K^*)\pf_b[K^*],
\end{equation}
with \(A\) some analytic function of the set of couplings $K$ of model
$a$,  and dual couplings $K^*$.  In principle, Eq. \eqref{class_d}
establishes an extremely broad  relationship that could be achieved
through many transformation schemes, including the standard one based on
taking the Fourier transform of individual Boltzmann weights 
\cite{Savit,malyshev}. However, it was discovered in Refs.
\cite{ours2,dualities} that low-temperature(strong
coupling)/high-temperature(weak coupling) dualities correspond to a
unitary equivalence of transfer matrices or operators,   $T_a$ and
$T_b$, 
\begin{equation}
T_b= \mathcal{U}_\d T_a \mathcal{U}_\d^\dagger,
\end{equation}
with $\mathcal{U}_\d$ a unitary operator.  This observation is extremely
insightful  because there is a  simple and systematic way to look for
unitary equivalences between physical operators, based on the notion of
{\it bond algebra}, or algebra of interactions \cite{ours2,dualities}.

The outline of this article is as follows. In Section \ref{revXY} we
define the classical XY model, and then in \ref{transferXY} {\it
establish its transfer operator}.  In Section \ref{hamiltonianXY} we
discuss the form of  the  exact one-dimensional quantum analogue whose
partition function is that  of the $D=2$ classical XY model with
coupling constants $K_1$ and $K_2$.  In the limit of large coupling
$K_{2}$ along columns, the  quantum model is the O(2) quantum rotor
model.     In Section \ref{XYtoints}, we establish the duality of the
$D=2$ XY model to a  solid-on-solid-like and also to a lattice
Coulomb gas-like models and,  moreover, determine   the disorder variables.
These dualities do not rely  on the Villain approximation scheme but are
{\it exact dualities} obtained by our bond-algebraic method
\cite{dualities}.

We next proceed to analyze in Section \ref{p-section}  the $p$-clock
model \cite{Potts}.  This model provides a particular  controlled limit
to the  XY model (the $p \to \infty$ limit).  We replicate the same
steps undertaken in the analysis of the XY model of Section
\ref{revXY},  but now the Weyl algebra \cite{Schwinger} and the theory
of circulant matrices  \cite{Davis} play a key role.  We construct in
Section \ref{tp}  its transfer matrix,  and proceed in
\ref{hamiltonianp} to establish the corresponding one-dimensional 
quantum Hamiltonian, that is not self-dual for $p\geq 5$.  We study the
dualities of these systems in Section \ref{dpclock}. The system is
exactly self-dual for $p=2,3,4$, and  becomes approximately self-dual
for $K_{2} \gg K_{1}$ when  $p \geq 5$.  In Section \ref{modify_p} we
introduce a variant of the classical $p$-clock model that is exactly
self-dual for all $p$.  We examine, in Section \ref{symmetriesp}, the
exact and emergent symmetries of these systems and, notably,  unveil the
U(1) symmetry that emerges  when $p \geq 5$. This  continuous  emergent
symmetry is responsible for the  existence of the intermediate critical
(massless) phase. 

Finally, in Section \ref{sec10} we utilize our previous findings to
better understand the phase diagram of the $p$-clock model. Here we
present an analytic expression for the  self-dual temperature $T^*$, an
important scale in the problem, and advance a Peierls argument.  We 
also introduce a topological invariant, that we call the discrete
winding number $k$, to unravel the  nature of the topological
excitations. Starting at $p \geq 5$ a new type of topological
excitation  appears with a non-zero value of $k$ that one may call
discrete vortex, and which is  responsible for the phase transition to a
disordered state. Domain-wall topological excitations are  key at low-temperatures
and their energy cost depends on $p$, and on the relative  spin
configurations (except for $2 \leq p \leq 4$). By using both duality and
energy-versus-entropy  balance considerations we show that the
transition from the critical to the disordered phase  scales as
$T^{(2)}\sim {\cal O}(1)$, while the one from the broken $\mathbb{Z}_p$
symmetry to  the critical phase goes as $T^{(1)}\sim 1/p^2$ for large
$p$. We derive a correlation inequality, i.e., show that the two-point 
correlation function $G$ is a monotonically decreasing function of 
temperature, allowing us to prove that,  for $p \geq 5$, the intermediate 
phase is  critical (massless) with decaying power-law correlations.
The appendices provide technical  developments including a duality 
of the XY model to $q$-deformed bosons,  illustrating the key physical 
difference between compact and non-compact degrees of freedom.

\section{The XY model: A paradigm of BKT phenomenology}
\label{revXY} 

\begin{figure}[thb]
\begin{center}
\includegraphics[angle=0,width=12cm]{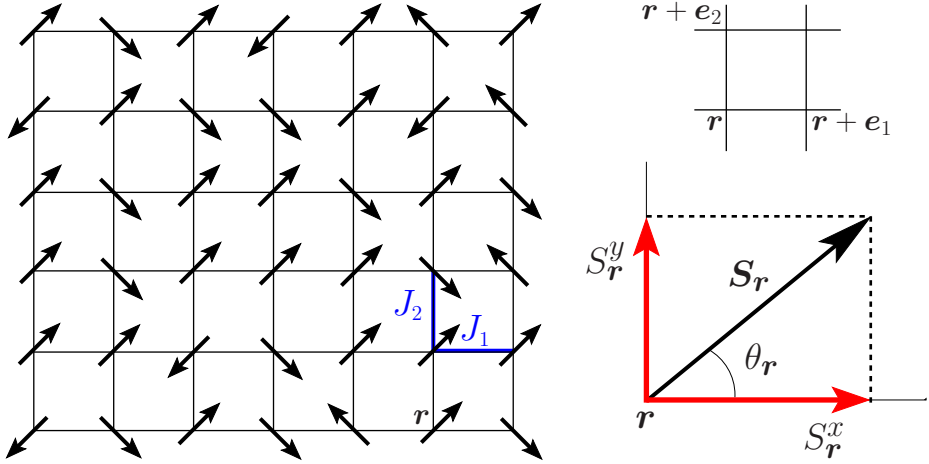}
\end{center}
\caption{The two-dimensional classical XY model.  On each vertex
$\bm{r}=i \, \i + j \, \j$ of the square lattice there is  a classical
spin $\bm{S}_{\bm{r}}=S^x_{\bm{r}} \, \bm{e}_{1} + S^y_{\bm{r}} \,
\bm{e}_{2} $ of  magnitude 1, i.e., $\bm{S}_{\bm{r}} \cdot
\bm{S}_{\bm{r}}=1$, and  $S^x_{\bm{r}} =\cos \theta_{\bm{r}}$,
$S^y_{\bm{r}} = \sin \theta_{\bm{r}}$ with $\theta_{\bm{r}} \in
[0,2\pi)$.  Nearest neighbor spins interact with an exchange constant
value $J_1$ or $J_2$ depending on  the spatial direction. }
\label{fig_XY}
\end{figure}

The $D=2$ classical XY model is the paradigmatic example  of a system
displaying a BKT transition at a finite temperature $T^{(2)}_{\sf
BKT}>0$. This  model, also known as planar rotator or planar O(2),  
consists of an $N\times N$ array of classical two-component spins
$\bm{S}_\r$ located at the  vertices $\r=i \, \i + j \, \j$ ($i,j$ being
integers) of  a square lattice with unit vectors $\bm{e}_\mu$, 
$\mu=1,2$, as indicated in Fig. \ref{fig_XY}. Its partition function is 
\begin{equation}
\pf_{\sf XY}[K_\mu,\bm{h}]=\sum_{\{\bm{S}_\r\}}\ \exp\left[\sum_{\r}
\left ( \sum_{\mu=1,2} K_\mu \, \bm{S}_{\r+\bm{e_\mu}}\! \cdot \bm{S}_\r
+ \bm{h} \cdot \bm{S}_\r \right ) \right],
\label{classicalXY}
\end{equation}
where the spin $\bm{S}_\r=S_\r^x \, \bm{e}_1+ S_\r^y \, \bm{e}_2$, the
coupling $K_\mu = \beta J_\mu$ is the product of the inverse
temperature  $\beta=1/k_B T$ and the exchange coupling $J_\mu$, and
$\bm{h}$ is a temperature-rescaled external magnetic field.  Fixing the
magnitude of the spin variable $\bm{S}^2_\r$ to one allows us  to
re-write the partition function as 
\begin{equation}
\pf_{\sf XY}[K_\mu,{h}]=\sum_{\{\theta_\r\}}\ \exp\left[\sum_{\r}\left
(\sum_{\mu=1,2} K_\mu \, \cos(\theta_{\r+\bm{e_\mu}}-\theta_\r) + h \cos
\theta_\r  \right)\right] ,
\label{classicalXYtheta}
\end{equation}
where the continuous angle variables $\theta_\r$  take values in the
interval $\theta_\r =\theta_{i,j} \in [0,2\pi)$,  i.e., it is a compact
variable. We assume, without loss of generality,  that  $\bm{h}=h \,
\bm{e}_1$. The sum over  configurations represents an integral
\begin{equation}
\sum_{\{\theta_\r\}}=\int_0^{2\pi}\prod_\r d\theta_\r .
\end{equation}

In the remainder of this article, we will concentrate on the case with
zero external magnetic field,  i.e., $h=0$. The XY model displays a
(global) continuous U(1) symmetry, which  amounts to the invariance of
the model under a simultaneous rotation of every spin in the  lattice by
the same angle.  In low dimensional systems with continuous symmetries,
such as the XY model above, long-range order  is more fragile.  Thermal
fluctuations may induce instabilities with the end result that 
long-range order is actually non-existent  in $D\leq 2$ dimensions. 
Spin-wave excitations are responsible for destroying such an order. The
{\em Mermin-Wagner theorem} formalizes this qualitative picture.  In the
context of the $D=2$ XY model, the theorem  states that this system 
does not display  spontaneous magnetization at finite temperatures. A
common physical mechanism behind the formal proofs  for both classical
and quantum versions of the XY model, can be found in Ref. \cite{NO}.

A phase transition is said to occur whenever a thermodynamic function of
the  system under study displays a non-analyticity. The latter may occur
even when the ground state is unique,  so that there is {\it no}
spontaneous symmetry breakdown. This is the case for the  $D=2$ XY
model, that is  known to have a special phase transition at a finite,
non-zero  temperature $T^{(2)}_{\sf BKT}$. This  BKT  {\it transition}
is  characterized by an essential singularity in the free energy and
correlation length at $T^{(2)}_{\sf BKT}$. If $T>T^{(2)}_{\sf BKT}$, the
correlators of the XY model decay exponentially with distance, as is 
typical of a disordered, paramagnetic phase. In the low-temperature
phase,  $T<T^{(2)}_{\sf BKT}$, the  correlators decay algebraically with
distance, just as if every temperature below  $T^{(2)}_{\sf BKT}$
represented an ordinary critical point. The fact that   this power-law
behaviour extends over the finite temperature range  $0<T<T^{(2)}_{\sf
BKT}$, without long-range order, is known as  {\it quasi-long-range
order}.

\subsection{A transfer operator for the  {\rm \bf XY} model}
\label{transferXY}

In this section we set up a transfer operator for the XY model, in
preparation for the detailed study of its duality properties  and
symmetries.  We assume open boundary conditions in the \(\i\)-direction
and periodic ones  in the \(\j\)-direction. 

Since we are considering the XY model on a square lattice of  size
\(N\times N\) we will need the operators 
\begin{equation}\label{opsXY}
L_{z,i}=-\im\frac{\partial}{\partial\theta_i},\ \ \mbox{ and }   \ \ \
e^{\pm \im\hat{\theta}_i}, \ \ i =1,2,\cdots, N ,
\end{equation}
satisfying the following commutation relations
\begin{equation}\label{h_circle}
[L_{z,i},\ e^{\pm \im\hat{\theta_j}}]=\pm\delta_{i,j} e^{\pm
\im\hat{\theta}_j}.
\end{equation}
The eigenstates of the unitary operators \(e^{\pm \im\hat{\theta}_i}\),
\begin{equation}
e^{\pm \im\hat{\theta}_i} \, |\theta_i\rangle=e^{\pm \im{\theta}_i} \,
|\theta_i\rangle, \ \ \ \ \theta_i \in [0,2\pi),
\end{equation} 
satisfy
\begin{equation}\label{circle_kets}
\langle\theta_i'|\theta_i\rangle=\delta(\theta_i'-\theta_i),\ \ \ \ \ \
\int_0^{2\pi}d\theta_i\ |\theta_i\rangle\langle\theta_i|=\mathds{1}.
\end{equation}
The plane wave eigenstates \(|n_i\rangle\) of \(L_{z,i}\) form an
orthonormal basis of the Hilbert space of square integrable functions 
on the cirle \(\mathcal{L}^2({\rm U(1)})=\mathcal{H}_i\), and are
related to the states \(|\theta_i\rangle\) via
\(\langle\theta_i|n_i\rangle= e^{\im \theta_i n_i}/\sqrt{2\pi}\). 

On the one hand, \(e^{\pm \im \hat{\theta}_i}\)  represent position
operators for the spin at site \(i\),  since their simultaneous
eigenstates \(|\theta_i\rangle\) specify one, and only one point on the
unit circle. On the other hand,  $L_{z,i}$ represents their canonically
conjugate momentum, the infinitesimal generator of  translations
\begin{equation}\label{circle_shift}
e^{-\im\delta L_{z,i}}|\theta_i\rangle=|\theta_i+\delta\rangle.
\end{equation}
This last equation follows from Eq. \eqref{h_circle}, since
\begin{equation}\label{circle_translation}
e^{\im\delta L_{z,i}}e^{\pm \im\hat{\theta}_i}e^{-\im\delta L_{z,i}}=
e^{\pm \im(\hat{\theta}_i+\delta)}.
\end{equation}
The product states,
\begin{equation}\label{basis}
|\theta\rangle=\bigotimes_{i}\ |\theta_{i}\rangle,
\end{equation}
that are simultaneous eigenstates of all the position operators 
\(e^{\pm \im \hat{\theta}_i}\), form an orthonormal basis of  the total
Hilbert space $\mathcal{H}=\bigotimes_{i}\mathcal{H}_i$.

We have now all the ingredients needed to write down a transfer operator
for  the XY model. Consider for concreteness the following row-to-row
($j$ to $j+1$) matrix elements of the transfer operator 
\begin{equation}\label{t2}
\langle\theta' \vert T_2|\theta \rangle= \exp\left[\sum_{i=1}^N
K_2\cos(\theta_{i,j+1}-\theta_{i,j})\right],
\end{equation}
and the diagonal operator 
\begin{equation}\label{t1}
T_1|\theta\rangle= \exp\left[\sum_{i=1}^{N-1}
K_1\cos(\theta_{i+1,j}-\theta_{i,j})\right] |\theta\rangle.
\end{equation}
Both matrices are defined in the basis of states introduced in Eq.
\eqref{basis}. It is straightforward to check that if we set \(T_{\sf
XY} \equiv T_2T_1\), then
\begin{equation}\label{XYt}
\tr [T_{\sf XY}^N]=\pf_{\sf XY}[K_\mu, h=0],
\end{equation} 
recovers the partition function for the XY model of Eq.
\eqref{classicalXY},  provided that we set the external magnetic field
\(h\) to zero. 
 
Next, we  rewrite the operators \(T_1,\ T_2\) in terms of the operators
introduced in  Eq. \eqref{h_circle}.  The result reads
\begin{equation}\label{t1t2}
T_1=\prod_{i=1}^{N-1}\ e^{K_1\cos(\hat{\theta}_{i+1}-\hat{\theta}_i)},\
\ \ \ \ \ T_2=\prod_{i=1}^{N}\ \int_0^{2\pi}d\theta \
e^{K_2\cos\theta}e^{-\im\theta L_{z,i}},
\end{equation} 
as can be checked by taking matrix elements of \(T_2T_1\) in the
basis of Eq. \eqref{basis}.  Notice that \(T_1\) factors into a product
of two-body operators that involves only nearest neighbors, while
\(T_2\) factors into a product of one-body operators. This important
simplification is a direct reflection of locality.

The relevant symmetries of the classical XY model translate into unitary
transformations that commute with \(T_{\sf XY}\equiv T_2T_1\). Besides
the obvious geometrical symmetries of the lattice, \(T_{\sf XY}\)
commute with two operators that represent internal, global symmetries.
The continuous global U(1) symmetry under global rotations of the
classical  spin direction  $\theta_\r\rightarrow\theta_\r+\alpha,\
\forall\r$, guarantees that   $[L_z,\ T_{\sf XY}]=0$, where
\(L_z=\sum_{i=1}^N\ L_{z,i}\) is the  total angular momentum. There is
also a discrete symmetry \({\mathcal{C}_0}=\prod_{i=1}^{N} {C}_{0i}\)  
that is alluded to, somewhat inaccurately, as ``charge conjugation'' 
\cite{Henkel}. The operator \({C}_{0i}\) acts on position eigenstates as
\({C}_{0i}|\theta_i\rangle=|2\pi-\theta_i\rangle\), and on angular
momentum eigenstates as \({C}_{0i}|n_i\rangle=| -n_i\rangle\). Thus,
\({C}_{0i}^2=\mathds{1}\) and \(C_{0i}^\dagger={C}_{0i}\). Since
\({\mathcal{C}_0}\) does not commute but rather anticommutes with
\(L_z\), we notice that the full group of internal symmetries of the XY
model is non-Abelian.

\subsection{Hamiltonian form of the {\rm \bf XY} model}
\label{hamiltonianXY}

Models that can be written in terms of an Hermitian transfer matrix or
operator,  such as the XY model, can be translated into 
quantum-mechanical problems \cite{NO} by simply defining a quantum 
Hamiltonian according to 
\begin{equation}
H_{\sf XY}=-\ln(T_{\sf XY}),\ \ \ \ \mbox{or equivalently},\ \ \ \ 
T_{\sf XY}=e^{-H_{\sf XY}}.
\end{equation}
While this is a powerful tool, often used in the literature, its actual
value is diminished by the technical problem of computing \(\ln(T_{\sf
XY})\) and, perhaps more importantly, because  \(H_{\sf XY}\) turns out
to be a highly non-local operator. The standard way out of this
difficulty is to make approximations  that solve both of these problems.
The qualitative picture that emerges is  intuitively appealing but
problematic if the approximations are not  reasonably controlled.

We will not  determine \(H_{\sf XY}\) in closed form, but rather we will
compute 
\begin{equation}
H_\mu=-\ln T_\mu,\ \ \ \ \ \ \mu=1,2.
\end{equation}
in closed form, and then exploit the Baker-Campbell-Haussdorf (BCH)
formula to obtain the expansion 
\begin{equation}
H\equiv-\ln(e^{-H_2}e^{-H_1})=H_1+H_2+[H_1,H_2]/2+\cdots.
\end{equation}
Since \(T_{\sf XY}=T_2T_1\), as defined in the previous section, is not
Hermitian, we have to set  \(H_{\sf XY}=(H+H^\dagger)/2\). This only
affects terms quadratic and higher order in the commutators. We will
also study the conditions for the non-diagonal (kinetic) part of the 
Hamiltonian \(H_2\) to reduce to the intuitively appealing form
\(H_2\propto \sum_i\ L_{z,i}^2/2\).

Referring back to the operator form of \(T_1,\ T_2\), Eq. \eqref{t1t2},
we see that it is straightforward to compute \(H_1\). Since \(T_1\) is
already in diagonal form, we obtain 
\begin{equation}
H_1=-\sum_{i=1}^{N-1} K_1\cos(\hat{\theta}_{i+1}-\hat{\theta}_{i}).
\end{equation}
On the other hand, $T_2$ is not diagonal. Computing \(H_2\) is further
simplified by the fact that \(T_2\) factors into the product of \(N\)
commuting one-body operators.  It follows that \(H_2=\sum_{i=1}^N\
H_{2,i}\), with 
\begin{equation}\label{matrix_el}
e^{-H_{2,i}}=\int_0^{2\pi}d\theta\ e^{K_2\cos \theta}e^{-\im\theta
L_{z,i}}.
\end{equation}
Next we notice that, since  \(e^{-\im\theta_1 L_{z,i}}e^{-\im\theta_2
L_{z,i}}=e^{-\im(\theta_1+ \theta_2) L_{z,i}}\), \(H_{2,i}\) should be
of the form (see  \ref{appF})
\begin{equation}
H_{2,i}=-\int_0^{2\pi}d\theta\ a_{K_2}(\theta)e^{-\im \theta L_{z,i}}.
\label{oneb_Ak2}
\end{equation}
By combining this expression with Eq. \eqref{matrix_el}, we get an 
equality between functions of \(L_{z,i}\) that can be evaluated
in that operator's diagonal basis. The result is an infinite set of 
equations
\begin{equation}\label{h2xy}
\int_0^{2\pi}d\theta  \ a_{K_2}(\theta)e^{-\im \theta n}= \ln ( 2\pi
I_n(K_2)),\ \ \ \ \ \ n\in\mathbb{Z},
\end{equation}
that one can use to determine \(a_{K_2}(\theta)\) by taking the Fourier
transform of the left-hand side. The modified Bessel functions of the
first kind and of  integer order \(n\), \(I_n(K_2)\), satisfy
\cite{bessel}
\begin{equation}\label{In}
e^{K_2\cos\theta}=\sum_{n\in \mathbb{Z}} I_n(K_2)e^{\im \theta
n}=I_0(K_2)+ 2 \sum_{n=1}^\infty I_n(K_2) \cos (\theta n),
\end{equation} 
after using the relation $I_{-n}(K)=I_n(K)$ \cite{bessel}. 

It follows from Eq. \eqref{h2xy} that \(a_{K_2}(2\pi-\theta)=
a_{K_2}(\theta)\), so that we can make the substitution \(e^{-\im \theta
L_{z,i}}\rightarrow \cos(\theta L_{z,i})\) in Eq. \eqref{oneb_Ak2}.  
Then, we can Taylor-expand the cosine function to get 
\begin{equation}\label{exp}
H_{2,i}=-\sum_{m=0}^\infty\ a_m(K_2)L_{z,i}^{2m},
%\ \ \ \ \ 
%a_m(K_2)=\frac{(-1)^m}{(2m)!}\int_0^{2\pi}d\theta\
%\theta^{2m}a_{K_2}(\theta).
\end{equation}
with \(a_m(K_2)=(-1)^m\int_0^{2\pi}d\theta\ \theta^{2m}a_{K_2}(\theta)/
(2m)!\).  This equation provides a very convenient representation of
\(H_{2,i}\),  especially if we are allowed to discard terms beyond
\(m=1\). To investigate this possibility, consider  Eqs.
\eqref{matrix_el} and \eqref{exp}, and  evaluate those expressions in
the \(L_{z,i}\)'s diagonal basis to get
\begin{equation}\label{couplings_hXY}
\sum_{m=0}^\infty\ a_m(K_2) \, n^{2m}=\ln(2\pi I_n(K_2)).
\end{equation}
For large $K_2$, the functions \(I_n(K_2)\) have  the following
asymptotic expansion
\cite{bessel}
\begin{equation}
I_n(K_2)\thicksim \frac{e^{K_2}}{\sqrt{2\pi K_2}}\sum_{m=0}^\infty \
(-1)^m \ \frac{c_m(n)}{K_2^m}, 
\label{asympBess}
\end{equation}
%(the notation \(f\thicksim g\) means that \(f(x)/g(x)\rightarrow 1\) 
%as \(x\rightarrow \infty\)), 
with
\begin{equation}
c_0(n)=1,\ \ \ \
c_m(n)=\frac{(4n^2-1)(4n^2-3^2)\cdots(4n^2-(2m-1)^2)}{m! \, 8^m}, m\geq
1.
\end{equation}
Notice that this can be rearranged into an expansion in \(n^2\) that 
can be compared to the left-hand side of Eq. \eqref{couplings_hXY}.
Keeping, for each \(m\), only the leading order in \(1/K_2\), and
expanding the logarithm accordingly ($\ln (1+x) \sim x$), we obtain 
\begin{equation}
\ln(2\pi I_n(K_2))\thicksim \ln \left ( \sqrt{\frac{2\pi}{K_2}} \,
e^{K_2}\right )+\sum_{m=1}^\infty \ (-1)^m \frac{n^{2m}}{2^m m! K_2^m}.
\end{equation}
Comparing with Eq. \eqref{couplings_hXY}, 
\begin{equation}
a_0(K_2)\thicksim \ln  \left ( \sqrt{\frac{2\pi}{K_2}} \, e^{K_2} \right
) ,\ \ \ \ a_m(K_2)\thicksim \frac{(-1)^m}{2^m m! K_2^m},\ \ \ m\geq 1 ,
\end{equation}
so that \(a_{m+1}/a_{m}\thicksim -1/(2(m+1)K_2)\). In summary, in the 
large \(K_2\) limit, 
\begin{equation}
H_{\sf XY}\approx -N a_0(K_2)+\sum_{i=1}^N\ \frac{1}{2K_2}L_{z,i}^{2} - 
\sum_{i=1}^{N-1} K_1\cos(\hat{\theta}_{i+1}-\hat{\theta}_{i}),
\label{approxhxy}
\end{equation}
which is the one-dimensional O(2) quantum rotor model. 

\subsection{Duality of the  {\rm \bf XY} model without the Villain
approximation}
\label{XYtoints}

We now exploit the detailed understanding we have gained on the exact
operator structure of the XY model to look for  its dual
representations. The standard approach to the dualities of the XY model
starts by replacing it  with the Villain model (see  \ref{villain}),
then mapping the latter to the solid-on-solid (SoS) model, and finally
mapping the SoS model to a lattice Coulomb  gas \cite{NO}. In contrast,
in this section we establish directly {\it exact}  dual representations
of the XY model.  In  \ref{xyqbosons}, we establish a duality to  a
$q$-deformed boson Hamiltonian which illustrates the fact that 
non-canonical bosons need to emerge because of the compact nature of 
the degrees of freedom of the XY model.

Our methodology starts with the transfer  operators \(T_1,\ T_2\) introduced
in Eq. \eqref{t1t2}. The algebra of interactions, or {\it bond algebra}
in the language of  Refs. \cite{ours1,dualities,ours2,ours0}, underlies
their basic structure.  In this case, this is the von Neumann algebra
\(\mathcal{A}_{\sf XY}\) generated by the bonds 
\begin{eqnarray}\label{bondsXY}
&&L_{z,1},\ \ \ L_{z,i},\ \ \ 
e^{\pm\im(\hat{\theta}_{i}-\hat{\theta}_{i-1})},\ \ \ \ \ \ i=2,\cdots,
N. \nonumber
\end{eqnarray} 
Notice that \(T_1,\ T_2\in\mathcal{A}_{\sf XY}\), since these operators
are expressible as sums of products of the bonds listed in  Eq.
\eqref{bondsXY}.  \(\mathcal{A}_{\sf XY}\) reflects the  interactions
present in the XY model and is at the same time easy to characterize in
terms of relations. Then we can look  for other dual realizations
\(\mathcal{A}^D_{\sf XY}\)  that are isomorphic images of
\(\mathcal{A}_{\sf XY}\). By the general properties of von Neumann
algebras, it must be that  \(\mathcal{A}^D_{\sf
XY}=\mathcal{U}_\d\mathcal{A}_{\sf XY}\mathcal{U}_\d^\dagger\), with 
$\mathcal{U}_\d$ unitary, provided both algebras act on state Hilbert
spaces of the same dimensionality.  This is all we need to establish a
duality for the XY model. The dual partition function is determined from
the dual transfer operator \(T^D_{\sf XY}=\mathcal{U}_\d T_{\sf XY}
\mathcal{U}_\d^\dagger\).

\begin{figure}
\centering
\includegraphics[angle=0, width=.7\columnwidth]{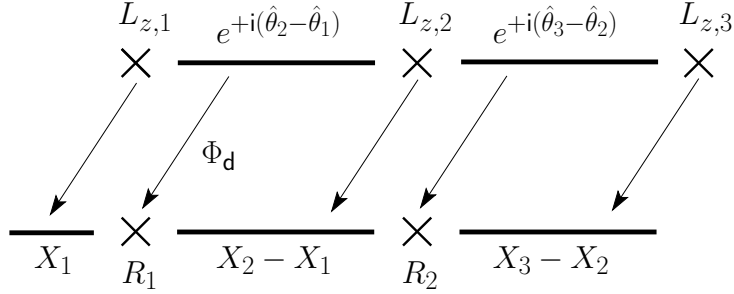}
\caption{The bond algebra isomorphism \(\Phi_\d(\cdot)=
\mathcal{U}_\d\cdot\mathcal{U}_\d^\dagger\)  defined in Eq.
\eqref{dXY_integers}, illustrated for \(N=3\) sites.}
\label{dXY_int}
\end{figure}

The goal of this section is to look for a dual representation of the  XY
model in terms of integer-valued degrees of freedom, so we can expect
the dual bond algebra \(\mathcal{A}^D_{\sf XY}\) to act  on the state
space \(\bigotimes_{i=1}^{N}\mathcal{L}^2(\mathbb{Z})\). Let us
introduce the states \(|n\rangle=\bigotimes_{i=1}^{N}|n_i\rangle\),
and the operators $X_i$, $R_i$
\begin{eqnarray}\label{integer_circle}
&&\hspace*{-0.7cm}X_i|n\rangle=n_i|n\rangle,\\
&&\hspace*{-0.7cm}R_i|n\rangle=|\cdots,n_{i-1},n_i-1,n_{i+1},\cdots\rangle,\ \ \ \ 
R^\dagger_i|n\rangle=|\cdots,n_{i-1},n_i+1,n_{i+1},\cdots\rangle,\nonumber
\end{eqnarray}
that satisfy the algebra
\begin{equation}
[X_i,\ R_j]=-\delta_{i,j}R_j,\ \ \ \ [X_i,\ R_j^\dagger]=
\delta_{i,j}R_j^\dagger,\ \ \ \ R_jR_j^\dagger=\mathds{1}.
\end{equation}
Then, the operators
\begin{eqnarray}
X_1,\ \ \ X_{i+1}-X_i,\ \ \ R_i,\ \ \ R_i^\dagger,
\ \ \ \ \ \  i=1,\cdots,N-1,
\end{eqnarray}
generate an isomorphic dual representation \(\mathcal{A}^D_{\sf XY}\) 
of the bond algebra of the XY model. The isomorphism \(\Phi_\d\) 
connecting the two bond algebras 
\begin{eqnarray}\label{dXY_integers}
L_{z,1}&\dual& X_1,\ \ \ \ L_{z,i}\dual X_{i}-X_{i-1},\ \ \ \ \ \
i=2,\cdots,N,\\
e^{+\im(\hat{\theta}_{i+1}-\hat{\theta}_{i})}&\dual& R_i,\ \ \ \ 
 e^{-\im(\hat{\theta}_{i+1}-\hat{\theta}_{i})}\dual R_i^\dagger
,\ \ \ \ \ \ i=1,\cdots, N-1,
\nonumber
\end{eqnarray}
is illustrated in Fig. \ref{dXY_int}, for \(N=3\) sites.
The resulting dual transfer operators are
\begin{eqnarray}
T_1^D&=&\prod_{i=1}^{N-1}e^{\frac{K_1}{2}(R_i+R_i^\dagger)}\ ,\\
T^D_2&=&\int_0^{2\pi}d\theta \, e^{K_2\cos\theta}e^{-\im X_1}\
\prod_{i=2}^{N}\int_0^{2\pi}d\theta \, e^{K_2\cos\theta}e^{-\im
(X_i-X_{i-1})}\ . \nonumber
\end{eqnarray}
The next and last step is to compute the dual partition function
\begin{equation}
\pf_{\sf XY}^D=\tr [(T_2^DT_1^D)^N]= \tr [(T_{\sf XY}^D)^N],
\end{equation}
in the product basis states of Eq. \eqref{integer_circle}.

\(T_2^D\) is already diagonal in that basis, leading simply to 
\begin{equation}
T_2^D|n\rangle=\exp\left[-V_{K_2}(n_{1,j})-\sum_{i=2}^{N}
V_{K_2}(n_{i,j}-n_{i-1,j})\right] |n\rangle,
\end{equation}
where 
\begin{equation}\label{vk}
V_K(n)=-\ln \int_0^{2\pi}d\theta \, e^{K\cos\theta}e^{-\im \theta n}=
-\ln (2\pi I_n(K)).
\end{equation}
The evaluation of \(\langle n'|T_1^D|n\rangle\) factors into 
the computation of the one-body matrix elements,
\begin{equation}
\langle n_{i}'|e^{\frac{K_1}{2}(R_i+R_i^\dagger)}|n_{i}\rangle.
\end{equation}
This is simplified by noticing that  the Fourier transform operator of
Eq. \eqref{circle_fourier} maps
\begin{equation}
F_i^\dagger R_iF_i=e^{-\im\hat{\theta}_i},\ \ \ \ \ 
F_i^\dagger R_i^\dagger F_i=e^{\im\hat{\theta}_i}
\end{equation}
thus putting the exponential in diagonal form. It follows that
\begin{equation}
\langle n_i'|e^{\frac{K_1}{2}(R_i+R_i^\dagger)}|n_i\rangle=2\pi
e^{-V_{K_1}(n_{i,j+1}-n_{i,j})}.
\end{equation}
%\(V_{K_1}\) is again the function of Eq. \eqref{vk}.
If we now put all the pieces together, we arrive at the conclusion  that
we have obtained the exact duality
\begin{equation}\label{xyintdual}
\frac{\pf_{\sf XY}[K_\mu]}{(2\pi)^N}=\sum_{\{n_{i,j}\}}\ e^{\left[
-\sum_{j=1}^N\sum_{i=1}^{N-1}\left(V_{K_2}(n_{i+1,j}-n_{i,j})+
V_{K_1}(n_{i,j+1}-n_{i,j})\right)+\sum_{j=1}^NV_{K_2}(n_{1,j})\right]}.
\end{equation}
This illustrates a typical characteristic of dualities: the coupling \(K_1\) (\(K_2\))
in the \(\i\)(\(\j\))-direction of the XY model regulates the
interaction in the orthogonal  \(\j\)(\(\i\))-direction of the dual
model. 

It is standard to argue that the Villain model is an  excellent
approximation to the XY model, specially at low temperatures. As shown
in  \ref{villain}, the Villain model is dual to the SoS model. Thus, it
must be that the SoS model is approximatly related to \(\pf_{\sf XY}^D\)
defined by the right-hand side of Eq. \eqref{xyintdual}, at least for
low  temperatures. Consider then  the limit of large \(K_1,\ K_2\)
(i.e., low temperatures).  We can then use the asymptotic expansion  of
Eq. \eqref{asympBess} to obtain an asymptotic form of the dual potential
of Eq. \eqref{vk},
\begin{equation}
V_K(n)\approx \frac{n^2}{2K}+c(K),\ \ \ \ K\rightarrow \infty ,
\end{equation}
where \(c(K)\) is independent of \(n\) and can be computed from Eqs. 
\eqref{vk} and \eqref{asympBess}. It follows that 
\begin{equation}
\pf_{\sf XY}^D\propto \sum_{\{n_{i,j}\}}\ e^{\left[
-\sum_{j=1}^N\sum_{i=1}^{N-1}(n_{i+1,j}-n_{i,j})^2/2K_2+
(n_{i,j+1}-n_{i,j})^2/2K_1+\sum_{j=1}^N(n_{1,j})^2/2K_2\right]} 
\end{equation}
to the same level of approximation. Thus we have recovered the  well
known result that the XY model at very low temperatures (strong
coupling) is well represented by the (approximately dual) SoS model at
very high temperatures (weak coupling). 

The action of the duality of Eq. \eqref{dXY_integers} can be extended to
act on the operator \(e^{\im\hat{\theta}_N}\) as 
\(\Phi_\d(e^{-\im\hat{\theta}_N})=R_N\). It follows that \(\Phi_\d\)
generates the following set of \(N\) dual variables (below we
distinguish a dual variable by an overtilde),
\begin{eqnarray}
\widetilde{e^{-\im\hat{\theta}_i}}&\equiv&\Phi_\d(e^{-\im\hat{\theta}_i})=
\prod_{m=i}^N\ R_m,\ \ \ \ i=1,\cdots,N\\
\widetilde{L_{z,1}}&\equiv&\Phi_\d(L_{z,1})=X_1,\ \ \ \
\widetilde{L_{z,i}}\equiv\Phi_\d(L_{z,i})=X_i-X_{i-1},\ \ \ \ 
i=2,\cdots,N.\nonumber
\end{eqnarray}
The dual variables satisfy the algebra of Eq. \eqref{h_circle},
confirming that \(\Phi_\d\) defines an algebra isomorphism.  Since this
is also the algebra of the variables \(R_i,\ R_i^\dagger,\ X_i\), we see
that
% and are
%represented on their Hilbert space as well. 
the dual variables  \(\widetilde{e^{\pm\im\hat{\theta}_i}}, \
\widetilde{L_{z,i}}\) afford an alternative representation of the
elementary degrees of freedom.  But what is their thermal behavior? This
crucial question can be answered easily because \(\Phi_\d\) amounts to a
unitary transformation. It follows that 
\begin{eqnarray}\label{dualcorrelator}
\langle e^{\im \theta_{m+r,n+s}}\ e^{-\im \theta_{m ,n }}\rangle
&=&\frac{\tr ( T_{\sf XY}^{(N-n-s)}\ e^{\im \hat{\theta}_{m+r}}\ 
T^{s}_{\sf XY}\ e^{\im \hat{\theta}_{m}} \ T_{\sf XY}^n)}{\pf_{\sf XY}}
\nonumber \\
&=&\frac{\tr ( T_{\sf XY}^{D (N-n-s)}\ \widetilde{e^{\im
\hat{\theta}_{m+r}}}\  T^{D s}_{\sf XY}\ \widetilde{e^{\im
\hat{\theta}_{m}}} \ T_{\sf XY}^n)}{\pf_{\sf XY}^D}   \nonumber \\
&=&\langle e^{\im
\tilde{\theta}_{m+r,n+s}}\ e^{-\im \tilde{\theta}_{m ,n }}\rangle,
\end{eqnarray}
that should be compared to 
\begin{equation}
\langle n_{m+r,n+s}\ n_{m ,n }\rangle =\frac{\tr ( T_{\sf XY}^{(N-n-s)}\
X_{m+r}\  T^{s}_{\sf XY}\ X_m \ T_{\sf XY}^n)}{\pf_{\sf XY}^D}.
\end{equation}
The classical dual variables \(e^{-\im \tilde{\theta}_{m ,n }}\) are
difficult to define directly, but they are well defined in the sense
that any correlator 
\begin{equation}
\langle e^{(-1)^{\sigma_1} \im \tilde{\theta}_{m_1,n_1}}\ 
e^{(-1)^{\sigma_2}\im \tilde{\theta}_{m_2,n_2}}\cdots\ 
e^{(-1)^{\sigma_N}\im \tilde{\theta}_{m_N,n_N}}\rangle,\ \ \ \
\sigma_i=0,1,
\end{equation}
in the ensemble \(\pf_{\sf XY}^D\) can be computed by a straightforward 
generalization of \eqref{dualcorrelator}.

The duality to a lattice Coulomb gas is of a very special nature
(Poisson duality)  and quite  different from every other duality
discussed in this article (or in literature on dualities in general).  
Its general features are discussed in Ref. \cite{dualities}. Here, we
briefly summarize the exact Coulomb gas-like dual model for the  exact
dual partition function \(\pf_{\sf XY}^D\) computed above. According to
Ref. \cite{dualities}, the lattice gas representation  of \(\pf_{\sf
XY}^D\) is defined by 
\begin{equation}
\pf_{\sf XY}^{DD}=\sum_{\{n_\r\}} e^{-\mathcal{E}^D\{n_\r\}},
\end{equation}
with interaction energy \(\mathcal{E}^D\) defined through the 
{\it global} Fourier transform 
\begin{equation}\label{pd}
e^{-\mathcal{E}^D\{n_\r\}}=\int\prod_\r dx_\r\ e^{\im 2\pi
\sum_\r n_\r x_\r}\ e^{\sum_\r\sum_{\mu=1,2} V_{K_\mu}(x_{\r+
\bm{e_\mu}}-x_\r)}.
\end{equation}
The interaction energy $\mathcal{E}^D$ can be determined in closed-form 
in the limit in which \(\pf_{\sf XY}^D\) reduces to the  SoS model, and
corresponds to the standard lattice Coulomb gas result \cite{NO}. We see
from Eq. \eqref{pd} that Poisson dualities are in general only of
practical use for models with Gaussian energy functionals.

\section{The $p$-clock model: A close relative of XY}
\label{p-section}

The $p$-clock model, also known as the vector Potts or $\mathbb{Z}_p$
model, represents a hierarchy of approximations to the XY model as a
function of the positive integer $p$, and is a test ground  for rich
critical behavior.  In $D=2$ dimensions, the configurations of the
classical $p$-clock model are  described by a set of \(p\) discretized
angles $\theta_{\r}$
\begin{equation}
\theta_\r=\frac{2\pi s_\r}{p},\ \ \ \ \ \ \ \ \ \ s_\r=0,1,\cdots,p-1.
\end{equation} 
The  partition function is given by
\begin{equation}\label{classicalVP}
\pf_p[K_\mu]=\sum_{\{\theta_\r\}}\ \exp\left[\sum_{\r}\sum_{\mu=1,2}\ 
K_\mu\cos(\theta_{\r+\bm{e_\mu}}-\theta_\r)\right].
\end{equation}
This is, in appearance, identical to the classical XY model ($h=0$),
except for the  essential fact that the degrees of freedom
$\{\theta_\r\}$ are  now discrete and countable. In the  large $p$ limit
($p \rightarrow \infty$), however, one  supposedly recovers the  XY
model. Since the \(p\) points \(e^{\im \theta_\r}\) close a
\(\mathbb{Z}_p\) subgroup of U(1) (the group of \(p\)'th roots of
unity), the \(p\)-clock model manages to provide an approximation to the
XY model that features a finite number of states per site, without
sacrificing the XY's natural group structure.
%The phase diagram of the \(p\)-clock model has been the subject of
%research for many years, but there are still open problems. For 
%example, the $p$-clock  model is known to exhibit a BKT phase
%transition,  {\it for sufficiently large $p$} \cite{frohlich},  yet to
%determine the smallest $p$ required for a  BKT transition is a problem
%currently under debate. 
Also like the XY model, the \(p\)-clock has interesting but  hard to
uncover duality properties \cite{Cardy}. We will address this problem 
by the same methods  applied to the XY model. 
%The starting point is to
%write a transfer matrix for \(\pf_p[K_\mu]\), and coach it in a
%convenient operator language. 
In fact, we will follow the discussion of the XY model as  closely as
possible, to highlight the connections between the two  models.

\subsection{A transfer matrix for the $p$-clock model}
\label{tp}

To introduce a transfer matrix for  the \(p\)-clock model, we need to
define a suitable Hilbert space and a set of basic kinematical
operators. Let \(\pf_p\) be defined on an \(N\times N\) square lattice
with open boundary conditions on the \(\i\)-direction and periodic ones
on the \(\j\)-direction. The states on each site \(i=1,\cdots, N\) of a
row can be described by orthonormal vectors
\begin{equation}\label{pstate}
|s_i\rangle,\ \ \ \ s_i=0,\cdots,p-1 ,
\end{equation}
such that \(s_i\) represents the discrete angle \(2\pi s_i/p\).  They
span the state space \(\mathcal{H}_{p,i}\) at site \(i\), so the  total
state space is just \(\mathcal{H}_p=\bigotimes_{i=1}^N
\mathcal{H}_{p,i}\). If we write \(|s\rangle\equiv
\bigotimes_{i=1}^N|s_i\rangle\) for elements of the product basis of
\(\mathcal{H}_{p}\), then we can define a matrix
\begin{equation}\label{t2vp}
\langle s' \vert T_2|s \rangle= \exp\left[\sum_{i=1}^N
K_2\cos(\frac{2\pi s_{i,j+1}}{p}-\frac{2\pi s_{i,j}}{p})\right],
\end{equation}
related to any pair of adjacent rows \(j,\ j+1\), and the diagonal matrix 
\begin{equation}\label{t1vp}
T_1|s\rangle= \exp\left[\sum_{i=1}^{N-1} K_1\cos(\frac{2\pi
s_{i+1,j}}{p}- \frac{2\pi s_{i,j}}{p})\right] |s\rangle.
\end{equation}
These definitions guarantee that \(\pf_p[K_\mu]=\tr[(T_p)^N]=\tr[(T_2T_1)^N]\).

The degrees of freedom of the $p$-clock model (at any one site of the
lattice) can take any value out of a discrete, equidistant subset of
points of the unit circle. 
%These points close a \(\mathbb{Z}_p\)  subgroup of the unit circle. 
To proceed in analogy to Section
\ref{transferXY}, we need to introduce position operators and their
conjugate momenta in this discrete setting. The formalism that emerges
was  used extensively by Schwinger in his work on the foundations of
quantum mechanics \cite{Schwinger}. In what follows, we consider only
one site (one degree of freedom), for the sake  of clarity. We will
consider all \(N\) sites again near the end of the section.

It is easy to restrict the position operators \(e^{\pm\im\hat{\theta}}\)
used for the XY model  to the subset of configurations available to a
{\it clock handle} in  the \(p\)-clock model. The result is the operator
\(U\) satisfying
\begin{equation}
U|s\rangle=\omega^{s}|s\rangle \ , \ \ \ \ \ s=0,\cdots,p-1,
\label{eigenvofU}
\end{equation}
with \(\omega\equiv e^{\im 2\pi/p}\) representing a \(p\)th root of
unity. The position operator \(U\) and its Hermitian-conjugate
\(U^\dagger\)  satisfy \(UU^\dagger=\mathds{1}=U^p\).   The momentum
operator \(V\) conjugate to \(U\) rotates any state counter-clockwise to
its {\it nearest-neighbor}
\begin{equation}
V|0\rangle=|p-1\rangle,\ \ \ \ 
V|1\rangle=|0\rangle,\ \ \ \ \cdots,\ \ \ \
V|p-1\rangle=|p-2\rangle .
\end{equation}
Momentum and position operators are represented, in (\(p\times p\)) 
matrix form, as
\begin{eqnarray} 
{V}=
\begin{pmatrix}
0& 1& 0& \cdots& 0\\
0& 0& 1& \cdots& 0\\
\vdots& \vdots& \vdots&      & \vdots \\
0& 0& 0& \cdots& 1\\
1& 0& 0& \cdots& 0\\
\end{pmatrix} \ , \mbox{ and } {U}={\sf
diag}(1,\omega,\omega^2,\cdots,\omega^{p-1}).
\label{VsandUs}
\end{eqnarray}
It follows that \(V^\dagger\) implements a clockwise rotation, and that
\(V V^\dagger=\mathds{1}=V^p\). 
%Also, $\det U=\det V=(-1)^{p-1}$.
The fundamental algebraic relation 
\begin{equation}\label{weyl_group_algebra}
VU=\omega UV
\end{equation}
follows directly from the definitions of \(U\) and \(V\). 

As is well known from quantum mechanics, the ordinary position operator
\(\hat{x}\) and its conjugate momentum operator \(\hat{p}\)  are related
by a Fourier transform \(\mathcal{F}\), a unitary transformation in the
space of wave functions. Essentially the  same holds for the operators
\(U,\ U^\dagger\) and \(V,\ V^\dagger\). The appropriate unitary
transformation in this context is the discrete Fourier transform \(F\),
that in matrix form reads
\begin{eqnarray}
{F}^{\dagger}=\frac{1}{\sqrt{p}}
\begin{pmatrix}
1& 1& 1& \cdots& 1\\
1& \omega& \omega^2& \cdots& \omega^{p-1}\\
1& \omega^2& \omega^4& \cdots& \omega^{2(p-1)}\\
\vdots& \vdots& \vdots&      &\vdots \\
1& \omega^{p-1}& \omega^{(p-1)2}& \cdots& \omega^{(p-1)(p-1)}
\end{pmatrix} .
\label{DFTmatrix}
\end{eqnarray}
This is also known as Schur matrix \cite{matveev}. 
It follows that \cite{Schwinger}
\begin{equation}\label{wga_aut}
FUF^\dagger=V^{\dagger},\ \ \ \ \ \ FVF^\dagger=U,
\end{equation}
and so the eigenvectors of $V$, $\tilde{s}=0,1,\cdots,p-1$,
\begin{eqnarray}
V\ket{\tilde{s}}= \omega^{\tilde{s}} \ket{\tilde{s}} \ , \ \mbox{ with } 
 \ket{\tilde{s}} = \frac{1}{\sqrt{p}} \sum_{s=0}^{p-1}\omega^{\tilde{s} s} 
 \ket{s} ,
\label{eigenvofV}
\end{eqnarray}
are easily determined via a Fourier transform of the eigenvectors of $U$.

In  the mathematical literature $V$ is known as the fundamental
circulant matrix. This is so as it generates the algebra of circulant
matrices \cite{Davis} (meaning that any circulant matrix \(C\) is of the
form \(C=\sum_{m=0}^{p-1}a_mV^m,\  a_m\in\mathbb{C}\)). Together, \(U\)
and \(V\) generate the full algebra of  $(p\times p)$ complex matrices
\cite{Schwinger},  that we continue to call the Weyl group algebra,  to
emphasize that we are working with a distinguished set of generators.
This shows that they constitute a convenient basis set of kinematic
operators, because we can write any other operator in terms of them. 

We need to re-introduce the row spatial index \(i\)  to apply the
technology just developed and rewrite  the transfer matrices of Eqs. 
\eqref{t1vp} and \eqref{t2vp} in operator form. In what follows, $U_i,\
U_i^\dagger,\ V_i,\ V_i^\dagger$,  for $i=1,\cdots, N$, will be our
basic set of operators. They {\it commute} at different sites,  satisfy
the relation \eqref{weyl_group_algebra} at any one site \(i\), and act
on the state space \(\mathcal{H}_p=\bigotimes_{i=1}^N
\mathcal{H}_{p,i}\). One then obtains
\begin{equation}\label{t2tvop}
T_1=\prod_{i=1}^{N-1}\ e^{\frac{K_1}{2} (U_{i+1}^\dagger
U_i+U_{i+1}U_i^\dagger)}, \ \ \ \ \ \  T_2=\prod_{i=1}^{N}\
\sum_{m=0}^{p-1}\ e^{K_2\cos(2\pi m/p)}V_i^{\dagger m}.
\end{equation}
This last expression for \(T_2\) follows from the fact that  \(\langle
s_i'|V_i^{\dagger m}|s_i\rangle=0\) unless \(s_i'-s_i \equiv m\) modulo
\(p\) (mod$(p)$). It should be compared to the analogous expression for
the continuum circle, Eq. \eqref{t1t2}.

\subsection{Hamiltonian form of the \(p\)-clock model}
\label{hamiltonianp}

In this section we compute the Hamiltonian form of the \(p\)-clock
model  following the strategy of  Section \ref{hamiltonianXY}. We start
by computing \(H_\mu=-\ln T_\mu,\ \mu=1,2\), with \(T_1,\ T_2\)  as
defined in Eq. \eqref{t2tvop}. 

Since \(T_1\) is diagonal, we can write 
\begin{eqnarray}
H_1=-\sum_{i=1}^{N-1} \frac{K_1}{2}(U_{i+1}^\dagger U_i+ U_{i}^\dagger
U_{i+1}).
\label{transfer_VPH0}
\end{eqnarray}
\(H_2=\sum_{i=1}^NH_{2,i}\), on the other hand, is not as easy to write
down.  \(H_{2,i}\) is defined as 
\begin{equation}
e^{-H_{2,i}}=\sum_{m=0}^{p-1}\ e^{K_2 \cos(2\pi m/p) }\ V_i^{\dagger m}.
\end{equation}
As explained in  \ref{appF}, we can solve this equation to obtain
\begin{equation}\label{non_diag_VP}
H_{2,i}=-\sum_{m=0}^{p-1}\ a_m(K_2) V_i^{\dagger m}\ ,
\end{equation}
with
\begin{equation}\label{manyhs}
a_m(K_2)=\frac{1}{p}\sum_{s=0}^{p-1}\ \cos(\frac{2\pi m s}{p})
\ln\left(\sum_{l=0}^{p-1} \ e^{K_2\cos(2\pi l/p)}\cos(\frac{2\pi l
s}{p}) \right)\ .
\end{equation}
Then, the  Hamiltonian \(H_p\) for the \(p\)-clock model follows
\begin{equation}\label{exacthp}
H_p= -\sum_{i=1}^{N-1}\frac{K_1}{2}(U_{i+1}^\dagger U_i+ U_{i}^\dagger
U_{i+1}) -\sum_{i=1}^{N}\sum_{m=0}^{p-1}\ a_m(K_2) V_i^{\dagger m},
\end{equation}
provided we truncate the BCH expansion of \(\ln T_p\)  to linear order
(see the discussion in Section \ref{hamiltonianXY}).  We notice for
future reference that the  discrete Fourier transform
$\hat{F}=\prod_{i=1}^N F_i$ maps  $H_p\rightarrow\hat{F}^\dagger
H\hat{F}=\tilde{H}_p$, with 
\begin{equation}
\tilde{H}_p= -\sum_{i=1}^{N-1}\frac{K_1}{2}(V_{i+1}^\dagger V_i+
V_{i}^\dagger V_{i+1}) -\sum_{i=1}^{N}\sum_{m=0}^{p-1}\ a_m(K_2)
U_i^{\dagger m}.
\end{equation}

As discussed in  \ref{appF}, the coefficients \(a_m(K_2)\) have simple
asymptotic forms in the limit \(K_2\rightarrow\infty\). The
corresponding approximation to \(H_p\) reads 
\begin{eqnarray}
H_p\approx &-&N a_0(K_2)-K_1 H_U - 2a_1(K_2) H_V  ,
\label{Hpself-dual}
\end{eqnarray}
with
\begin{eqnarray}
H_{U}=\frac{1}{2}(U_N+U_N^\dagger+\sum_{i=1}^{N-1}(U_{i+1}^\dagger U_i+
U_{i}^\dagger U_{i+1}) ) \ , \ 
H_{V}=\frac{1}{2}\sum_{i=1}^{N} (V_i+V_i^{\dagger}) ,
\label{Hpself-dual2}
\end{eqnarray}
and (see Eq. \eqref{plarge_K})
\begin{equation}
a_1(K_2)\approx e^{K_2(\cos(2\pi/p)-1)},\ \ \ \ a_0(K_2)\approx K_2.
\end{equation}
Equation \eqref{Hpself-dual}  shows a boundary term 
(\(-K_1(U_N+U_N^\dagger)/2\)) not present in Eq. \eqref{exacthp},
and that we include to make this approximation to \(H_p\) 
{\it exactly self-dual} \cite{dualities}.

The approximation made in going from Eq. \eqref{exacthp} to Eq.
\eqref{Hpself-dual}, that keeps only \(V_i^\dagger\) and
\(V_i^{\dagger(p-1)}=V_i\), is reminiscent of the one introduced in
Section \ref{hamiltonianXY} based on Eq. \eqref{couplings_hXY}, whereby
we  replaced the operator \(\cos(\theta L_{z,i})\) for the simpler
\(L_{z,i}^2\) in Eq. \eqref{approxhxy}. Indeed, the two aproximations
coincide  in the \(p\rightarrow \infty\) limit. A simple way to see this
is to notice that we can realize the operator \(V^\dagger_i\) directly
in the Hilbert space  of the XY model as \(V^\dagger_i\rightarrow
e^{-\im 2\pi L_{z,i}/p}\). Then, in the limit \(p\rightarrow \infty\),
\begin{equation}
V_i+V^\dagger_i\  \rightarrow \ 
e^{\im 2\pi L_{z,i}/p}+e^{-\im 2\pi L_{z,i}/p}\approx 
2-(2\pi/p)^2 L_{z,i}^2.
\end{equation}

\subsection{Dualities of the  $p$-clock model}
\label{dpclock}

\begin{figure}
\centering
\includegraphics[angle=0, width=.7\columnwidth]{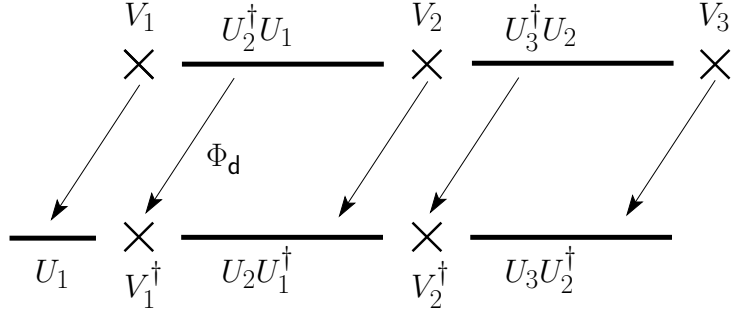}
\caption{The duality isomorphic mapping $\Phi_\d$ 
of Eq. \eqref{sdvp}, for \(N=3\) sites.}
\label{dvp}
\end{figure}

The dualities of the \(p\)-clock model appear as isomorphic
representations of the bond algebras associated to the transfer matrices
defined in Eq. \eqref{t2tvop}.  The bond algebra \(\mathcal{A}_p\)
generated by 
\begin{equation}\label{bondsvp}
V_1,\ V_1^\dagger,\ V_i,\ V_i^\dagger,\ U_{i}U_{i-1}^\dagger,\ 
U_{i}^\dagger U_{i-1},\ \ \ \ i=2,\cdots,N,
\end{equation}  
is simple to work with and adequate to our purposes. It has a dual 
(isomorphic) representation \(\mathcal{A}^D_p\) generated by the  same
bonds listed in Eq. \eqref{bondsvp}, except for \( V_N,\ V_N^\dagger\)
that have to be removed from the set of generators, and replaced by
\(U_1,\ U_1^\dagger\). The duality isomorphism \(\Phi_\d:\mathcal{A}_p
\rightarrow \mathcal{A}^D_p\) reads
\begin{eqnarray}\label{sdvp}
U_{i+1}^\dagger U_{i}&\dual& V_i^\dagger,\ \ \ \ \ \ 
\ \ \ \ \ \ \ \ \ \ \ \ \ \ \ \ \ \ \ \ \ \  \ \ \ \, i=1,\cdots, N-1,\\
V_1&\dual& U_1,\ \ \ \ V_i\dual U_{i-1}^\dagger U_{i},\ \ \ \ \ \
i=2,\cdots,N, \nonumber
\end{eqnarray}
together with the corresponding Hermitian-conjugate entries  (since it
must happen that \(\Phi_\d(\mathcal{O}^\dagger)=
\Phi_\d(\mathcal{O})^\dagger\)). \(\Phi_\d\) is illustrated in Fig.
\ref{dvp}, and should be compared with the duality of Section
\ref{XYtoints} for the XY model, illustrated in Fig. \ref{dXY_int}.

The dual form \(\pf_p^D=\tr[(T_2^DT_1^D)^N]\) of the \(p\)-clock model 
that follows from Eq. \eqref{sdvp} is defined by the dual transfer
matrices
\begin{eqnarray}
T_1^D&=&\prod_{i=1}^{N-1}e^{\frac{K_1}{2}(V_i+V_i^\dagger)}\ ,\\
T^D_2&=& \sum_{m=0}^{p-1}\ e^{K_2\cos(2\pi m/p)}U_1^{\dagger m}\ \times
\prod_{i=2}^{N}\ \sum_{m=0}^{p-1}\ 
e^{K_2\cos(2\pi m/p)}(U_{i}^\dagger U_{i-1})^{m}.
\nonumber
\end{eqnarray} 

Clearly, the $p$-clock model {\it is not self-dual} for arbitrary \(p\)
and arbitrary couplings. However, the model is approximately self-dual
in the extreme anisotropic limit with \(K_2 \gg K_1\), and it is exactly
self-dual for \(p=2,3,4\), and any coupling. We study these aspects of
the  $p$-clock model in the next  sections and in  \ref{p234}. 

We mention, in closing, that there is a   \(p\)-state model that
approximates the \(p\)-clock in the same sense  in which the Villain
model approximates the XY model. This $\mathbb{Z}_p$  Villain model
\cite{Elitzur} is exactly self-dual for any $p$, but is otherwise quite
different from the self-dual \(p\)-clock model to be  introduced next. 

\subsection{Self-dual classical $p$-clock model}
\label{modify_p}

In this section we introduce a classical model \(\pf_{\sf sdp}\)  that
we call the self-dual \(p\)-clock. It is  closely related to the
\(p\)-clock model (identical for $2 \leq p\leq 4$),  yet it is exactly
self-dual for any value of $p$, and has the  distinct advantage over the
$\mathbb{Z}_p$  Villain model \cite{Elitzur} that its transfer matrix is
remarkably simple.  To define the self-dual \(p\)-clock, we introduce
the transfer matrices 
\begin{equation}\label{t_sdpclock}
T_1[K_1]=e^{\frac{K_1}{2} (U_N+U_{N}^\dagger)}\ 
\prod_{i=1}^{N-1} e^{\frac{K_1}{2} (U_{i+1}^\dagger
U_i+U_{i+1}U_i^\dagger)}, \ \ \ \   T_2[a_0,a_1]=\prod_{i=1}^N\
e^{a_0+a_1(V_i+V_i^\dagger)}\ .
\end{equation}
Then, \(\pf_{\sf sdp}[a_0,a_1,K_1]=\tr [T_{\sf sdp}^N]\), with  \(T_{\sf
sdp}=T_2[K_2]T_1[a_0,a_1]\), and where \(a_0\) and \(a_1\) are free
parameters of the model to be determined, for  instance, by the
requirement that the approximation 
\begin{equation}\label{ttsdpapprox}
\sum_{m=0}^{p-1} e^{K_2 \cos{(2\pi m/p)}} V_i^{\dagger m} \approx 
e^{a_0+a_1(V_i+V_i^\dagger)},
\end{equation}
be as good as possible  for arbitrary \(K_2\). 
%This type of constraint on \(a_0\) and \(a_1\) is close in spirit 
%to the original work of Villain \cite{villain}.

\(\pf_{\sf sdp}\) is self-dual due to the existence of a  unitary 
transformation that maps
\begin{equation}\label{sdpct}
T_1[K_1]\rightarrow T_2[0,K_1],\ \ \ \ \ \ 
T_2[a_0,a_1] \rightarrow e^{Na_0}T_1[a_1],
\end{equation}
This fact results  from a bond-algebraic analysis, but we omit  the
details which can be found in Ref. \cite{dualities}. It follows from Eq.
\eqref{sdpct} that the self-dual line is specified by  \(K_1=2a_1\). The
next issue then is to understand the structure of \(\pf_{\sf sdp}\) in
terms of classical variables. On one hand, it is clear that the
interaction energy in the  \(\i\)-direction is still of the form
\(K_1\cos(2\pi(s_{i+1,j}-s_{i,j})/p)\). On the other hand,  the
interaction energy in the  \(\j\)-direction, \(u(s_{i,j}-s_{i,j+1})\),
is determined by the relation 
\begin{equation}
\sum_{m=0}^{p-1} e^{u(m)} V_i^{\dagger m}=e^{a_0+a_1(V_i+V_i^\dagger)}.
\end{equation}
Then, from Eq. \eqref{cos_bs_and_as}, 
\begin{equation}\label{sdu}
e^{u(m)}=\frac{1}{[p/2]}\sum_{s=0}^{[p/2]}\ \cos (\frac{2\pi ms}{p}) 
\ e^{a_0+2a_1\cos(2\pi  s/p)}
\end{equation}
(\([p/2]\) denotes the largest integer smaller than or equal to
\(p/2\)). With the interaction potential \(u\) defined in this way, 
\begin{equation}
\pf_{\sf sdp}[a_0,a_1,K_1]=\sum_{\{s_{i,j}\}}e^{\sum_{i,j}\left[
u(s_{i,j}-s_{i,j+1})+K_1\cos(2\pi(s_{i+1,j}-s_{i,j})/p)\right]}
\end{equation}
is exactly self-dual under the exchange \(2a_1\leftrightarrow K_1\).

One can see directly from Eq. \eqref{sdu} that \(u(p-m)=u(m)\).  It
follows that 
\begin{equation}
u(m)=
\sum_{r=0}^{[p/2]} K_{2,r} \cos (\frac{2\pi rm}{p}),
\end{equation}
with couplings 
\begin{equation}
K_{2,r}= a_0\delta_{r,0}+\frac{1}{[p/2]} \sum_{m=0}^{[p/2]}  \cos
(\frac{2\pi rm}{p})\ln\left[\frac{1}{[p/2]}\sum_{s=0}^{[p/2]}\  \cos
(\frac{2\pi ms}{p})e^{2a_1\cos(2\pi  s/p)}\right]
\end{equation}
determined by Eq. \eqref{sdu} and the orthogonality relation of Eq. 
\eqref{orthocos}. The point to notice is  that to make the \(p\)-clock
self-dual, we  need to add higher-order harmonics (terms \(\cos(2\pi
m(s_{i,j}-s_{i,j+1})/p)\), with \(m=2,\cdots,[p/2]\))  to the basic
cosine interaction. We show next that \(\pf_{\sf
sdp}\) becomes a very good approximation to the standard  \(p\)-clock
model in a suitable limit.

A comparison of the self-dual \(p\)-clock to the standard \(p\)-clock 
model shows that the latter has an approximate self-duality for $p \geq
5$.  As explained in \ref{appF}, in the limit in which \(K_2\) is large,
Eq. \eqref{ttsdpapprox} becomes almost exact, with  
\begin{equation}
a_0\approx K_2, \ \ \ \ \ \ a_1\approx e^{K_2(\cos\frac{2\pi}{p} -1)}
\end{equation}
(see Eq. \eqref{plarge_K}), so that 
\begin{eqnarray}
{\pf}_p[K_\mu]&\approx& e^{N^2K_2}\tr\left
[\left(e^{2a_1H_V}e^{K_1H_U} \right )^N \right ]\nonumber\\
&=&e^{N^2K_2}\tr\left [ \left( e^{2a_1H_U}e^{K_1H_V} 
\right )^N \right ] \nonumber\\
&\approx&e^{N^2(K_2-K_2^*)} {\pf}_p[K_\mu^*]
\end{eqnarray}
(see Eq. \eqref{Hpself-dual2}), with dual couplings
\begin{equation}
K_1^*= 2 e^{K_2(\cos\frac{2\pi}{p} -1)}\ ,\ \ \ \ \ \ \ 
K_2^*=\frac{\ln(K_1/2)}{\cos\frac{2\pi}{p} -1} .
\end{equation}

We emphasize that this approximate self-duality, in the extreme
anisotropic limit, is valid for {\it any} value of $p$. We  consider
{\it exact} self-dualities for the  particular cases $p=2,3,4$ in
\ref{p234}.

\subsection{Exact  and emergent symmetries of the $p$-clock model}
\label{symmetriesp}

{\it Non-Abelian, discrete symmetries}.
The representation of the transfer matrix \(T_p=T_2T_1\),  Eq.
\eqref{t2tvop}, is very convenient for understanding  the internal,
global symmetries of the \(p\)-clock model. It is apparent that the
model has an  Abelian \(\mathbb{Z}_p\) symmetry, but, as it turns out,
its full group of symmetries is considerably larger and non-Abelian,
provided \(p\geq3\).  To prove this we will show that there are two
Hermitian operators
\begin{equation}\label{polyhedral}
\mathcal{C}_0=\prod_{i=1}^N C_{0 i},\ \ \ \ \ \ 
\mathcal{C}_1=\prod_{i=1}^N C_{1 i},
\end{equation}
that commute with \(T_p\) and satisfy
\begin{equation}\label{polyhedral1}
\mathcal{C}_0^2=\mathcal{C}_1^2=(\mathcal{C}_0\mathcal{C}_1)^p=\mathds{1}.
\end{equation}
These relations show that, if  \(p\geq3\), \(\mathcal{C}_0\) and
\(\mathcal{C}_1\) generate a unitary representation of the so called
polyhedral group \(P(2,2,p)\)    \cite{rotman} of order \(2p\), and so
the group of internal symmetries of the  \(p\)-clock model is at least
as big as this non-Abelian group.  Notice that
\(\mathcal{C}_0\mathcal{C}_1\equiv\hat{Q}\), known as  $\mathbb{Z}_p$
charge,  generates a  \(\mathbb{Z}_p\) subgroup of \(P(2,2,p)\). This is
the standard Abelian symmetry  of the \(p\)-clock model that  gets
broken in  the low-temperature ordered phase (see Section \ref{sec10}).
It becomes a  U(1) symmetry in the limit $p \rightarrow \infty$, and
corresponds  to the usual continuous symmetry of the classical XY model
$\theta_\r\rightarrow  \theta_\r+\alpha$, with $\alpha$ an arbitrary
real number.

As \(\mathcal{C}_0\) and \(\mathcal{C}_1\) are products of one-site
operators, let us focus on a single site \(i\) for now. We define the
operators \(C_{0 i}\) and \(C_{1 i}\) by specifying their action on the
basis of Eq. \eqref{pstate},
\begin{equation}
C_{0 i}|s_i\rangle=|-s_i\rangle,\ \ \ \ C_{1 i}|s_i\rangle=|1-s_i\rangle,\ \ \ \
s_i=0,\cdots,p-1.
\end{equation}
The arithmetic in these definitions is modular, mod(\(p\)).  For
example, if \(p=5\), then \( C_{0 i}|0\rangle=|-0\rangle=|0\rangle\), \(
C_{0 i}|1\rangle=|-1\rangle=|4\rangle\), and so on. Keeping this in
mind, one can check that  \(\mathcal{C}_0\) and \(\mathcal{C}_1\) are
Hermitian,  \(C_{0 i}^\dagger=C_{0i}\), \(C_{1 i}^\dagger=C_{1i}\), and
satisfy the  relations listed in Eq. \eqref{polyhedral1} for 
\(\mathcal{C}_0\) and \(\mathcal{C}_1\). In particular, as
\(C_{0i}C_{1i}=V_i\),  \((C_{0i}C_{1i})^p=\mathds{1}\). If \(p=2\), then
\({\cal C}_{0}=\mathds{1}\),   and \(\mathcal{C}_1\) generates the
Abelian \(\mathbb{Z}_2\)  symmetry of the Ising model which is different
from the non-Abelian $P(2,2,2)$.

A routine calculation shows the action of \(C_{0 i},\ C_{1 i}\) on the
discrete position \(U_i\) and momentum \(V_i\) operators,
\begin{eqnarray}
\mathcal{C}_0V_i\mathcal{C}_0&=&V_i^\dagger,\ \ \ \ \ \ \ \ \ \  
\mathcal{C}_1V_i\mathcal{C}_1= V_i^\dagger, \\
\mathcal{C}_0U_i\mathcal{C}_0&=&U_i^\dagger,\ \ \ \ \ \ \ \ \ \ 
\mathcal{C}_1U_i\mathcal{C}_1=\omega U_i^\dagger.
\end{eqnarray}
It is easy to check that  \(\mathcal{C}_0\) and \(\mathcal{C}_1\)
commute with \(T_p\).  The operator \(\mathcal{C}_0\) is known in the
literature as the ``charge-conjugation'' operator \cite{Henkel}.
However, as we alluded to earlier,  this is something of a misnomer.
Geometrically speaking, \(\mathcal{C}_0\) is the exact analogue of the
parity operator \(\mathcal{P}|x\rangle=|-x\rangle\) on the real line. 
In fact, ${\cal C}_{0}$ is related to the discrete Fourier transform as 
$\hat{F}^2=(\hat{F}^\dagger)^2={\cal C}_{0}$, just as its counterpart on
the real line \(\mathcal{F}\) is connected to the parity operator as 
\(\mathcal{F}^2=\mathcal{P}\).

We wish to emphasize that these non-Abelian symmetries are shared by a 
large number of classical and quantum \(p\)-state models besides the
\(p\)-clock, including the self-dual \(p\)-clock introduced in Section
\ref{modify_p}  and  the \(\mathbb{Z}_p\) Villain models. 

{\it Emergent {\rm U(1)} symmetry.} For $p \geq 5$ the {\it discrete}
charge symmetry $\hat{Q}$ gets enhanced  into a {\it continuous} U(1)
symmetry. In reality, this is not an exact symmetry it is an  {\it
emergent} one \cite{BO}, but it is essential to establish the
intermediate BKT critical (massless) phase (see  Section \ref{sec10}). Let us
derive this emergent symmetry.

Given the generators of the SU(2) algebra in the spin $S=(p-1)/2$
representation
\begin{eqnarray} 
{S^z}&=&
\begin{pmatrix}
\frac{p-1}{2}& 0& 0& \cdots& 0 & 0\\
0& \frac{p-3}{2}& 0& \cdots& 0 & 0\\
\vdots& \vdots& \vdots&      & & \vdots \\
0& 0& 0& \cdots&\frac{3-p}{2} & 0  \\
1& 0& 0& \cdots & 0 & \frac{1-p}{2}\\
\end{pmatrix}, \nonumber \\
S^+&=&\begin{pmatrix}
0& {\scriptstyle \sqrt{p-1}}& 0& \cdots& 0 & 0\\
0& 0& {\scriptstyle \sqrt{2(p-2)}}& \cdots& 0 & 0\\
\vdots& \vdots& \vdots&      & & \vdots \\
0& 0& 0& \cdots& {\scriptstyle \sqrt{2(p-2)}}& 0  \\
0& 0& 0& \cdots & 0 & {\scriptstyle \sqrt{p-1}}\\
\end{pmatrix} , \nonumber
\label{VsandUsandS}
\end{eqnarray}
and $S^-=(S^+)^\dagger$, one may study the transformation properties of 
the Weyl's group generators $U$ and $V$ under the U(1) mapping 
\begin{eqnarray} 
\mathcal{U}_\phi=e^{-\im \phi S^z} .
\end{eqnarray} 
Since $U=\omega^{\frac{p-1}{2}} \,  \mathcal{U}_{2\pi/p}$, it commutes
with  $\mathcal{U}_\phi$. The transformation of $V$ requires some
thinking: Let us  rewrite $V$, Eq. \eqref{VsandUs}, as the sum of two 
operators
\begin{eqnarray} 
V=\hat{V} + \hat{\Delta} \ ,  \mbox{ with } \hat{\Delta}=
\begin{pmatrix}
0& 0& 0& \cdots& 0\\
0& 0&0& \cdots& 0\\
\vdots& \vdots& \vdots&      & \vdots \\
0& 0& 0& \cdots& 0\\
1& 0& 0& \cdots& 0\\
\end{pmatrix} \ , 
\end{eqnarray} 
i.e., the matrix that has only a $1$ in the lower-left corner. Then,
\begin{eqnarray}
\mathcal{U}_\phi   \hat{V} \, \mathcal{U}_\phi^\dagger = e^{-\im \phi}
\hat{V} \ ,  \mbox{ and } \mathcal{U}_\phi  \, \hat{\Delta} \,
\mathcal{U}_\phi^\dagger  = e^{\im (p-1) \phi} \hat{\Delta} .
\end{eqnarray}
We are interested in analyzing how the transfer matrices $T_1$ and $T_2$
of  Eq. \eqref{t2tvop} transform under
$\hat{\mathcal{U}}_\phi=\prod_{i=1}^N\mathcal{U}_{\phi,i}$.  It is
indeed easier to  analyze the Fourier transform  transfer matrices,
$\tilde{T}_\mu=\hat{F}^\dagger T_\mu \hat{F}$,
\begin{eqnarray}
\tilde{T}_1=
\prod_{i=1}^{N-1}\ e^{\frac{K_1}{2} (V_{i+1}^\dagger
V_i+V_{i+1}V_i^\dagger)}, \ \  \tilde{T}_2= \prod_{i=1}^{N}\
\sum_{m=0}^{p-1}\ e^{K_2\cos(2\pi m/p)}U_i^{\dagger m}.
\end{eqnarray}
Clearly, $\hat{\mathcal{U}}_\phi$ commutes with $\tilde{T}_2$ but does
not with $\tilde{T}_1$  unless $\phi=2\pi/p$, which not surprisingly
corresponds to the (Fourier transform)  discrete $\hat{Q}$ symmetry.
However, $\hat{\mathcal{U}}_\phi$ is an  {\it exact continuous}
symmetry of the modified transfer matrix 
\begin{eqnarray}
\widehat{T}_1= \prod_{i=1}^{N-1}\ e^{\frac{K_1}{2}
(\hat{V}_{i+1}^\dagger  \hat{V}_i+\hat{V}_{i+1}\hat{V}_i^\dagger + 
\hat{\Delta}_{i+1}^\dagger
\hat{\Delta}_i+\hat{\Delta}_{i+1}\hat{\Delta}_i^\dagger)},
\end{eqnarray}
and becomes the usual U(1) symmetry of the XY model when $p \rightarrow
\infty$.  This emergent symmetry may allow for the construction of 
spin-wave excitations in the critical region. 
%by employing the twist operator $e^{\im \frac{2\pi}{N} \sum_{i=1}^N \, i \, S_i^z}$. 
Note that in the original transfer matrix representation $T_1, T_2$, 
the continuous emergent symmetry is represented by  $ \hat{F} \,
\hat{\mathcal{U}}_\phi \hat{F}^\dagger$. Moreover, it is an emergent
symmetry of  {\it both} $p$-clock and self-dual $p$-clock models.

\section{Phase diagram: {}From the \(p\)-clock to the {\rm \bf XY} model}
\label{sec10}

\begin{figure}[thb]
\begin{center}
\includegraphics[angle=0,width=12cm]{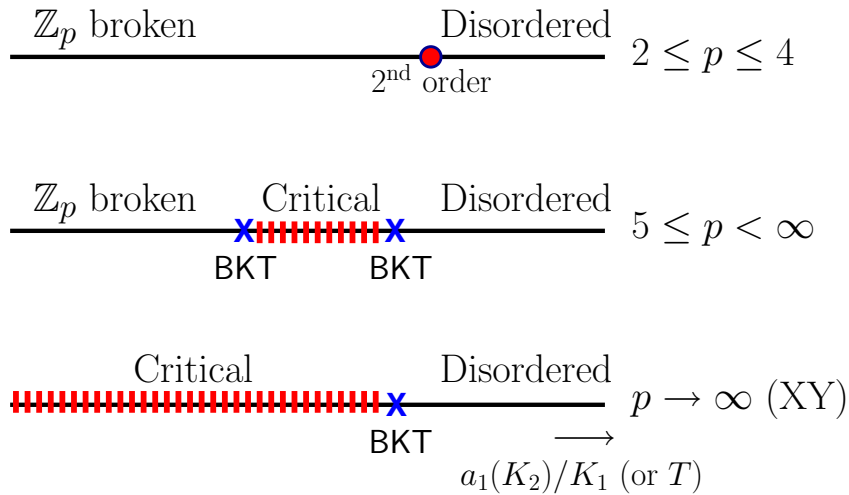}
\end{center}
\caption{Phase diagram of the $p$-clock model. For $p \geq 5$ there are
three phases, the  broken $\mathbb{Z}_p$ (low-temperature) phase
disappearing in the limit $p \rightarrow \infty$ (XY limit). A
transition is of BKT-type  whenever it is associated to an essential
singularity of the free energy. The critical phase is characterized by 
power-law correlations, i.e., quasi-long-range order, with non-universal
exponents.}
\label{fig_PD}
\end{figure}

We are thus left with the task of establishing the phase diagram of the 
$p$-clock model, the nature of its phase transitions and excitations,
and its behavior as \(p\rightarrow\infty\). One may argue that  the
phase structure of the model is well understood \cite{Elitzur}  (see
Fig. \ref{fig_PD}).  At very low temperatures,  there is a ferromagnetic
phase characterized by long-range order of the two-point, spin-spin, correlation
function $G(|\r - \r'|)= \langle \cos( \theta_\r-\theta_{\r'} )
\rangle$,  and the breakdown of the $\mathbb{Z}_p$ symmetry $\hat{Q}$.
At very high temperatures, the system is in a disordered phase with 
$G(|\r -\r'|)$ decaying as an exponential function of the distance.  For
$2 \leq p \leq 4$, these two phases  are separated by a continuous 
second-order phase transition of the Ising ($p=2,4$) or  Potts ($p=3$)
type.  (It is very easy to prove that the $p=4$ case  is {\it identical}
to two uncoupled $p=2$ Ising models \cite{class}, see  \ref{p234}.)  For
$p \gtrsim 5$  there is an additional  intermediate {\it critical} phase
separating the ferromagnetic from the  disordered phase.  It is
characterized by a power-law behavior of  $G(|\r - \r'|) \sim |\r -
\r'|^{-\eta}$  with a non-universal exponent $\eta$,  and by the absence
of symmetry breakdown and quasi-long-range order. In the $p \rightarrow
\infty$ limit the broken-symmetry phase disappears  as one recovers the
\(D=2\) XY model with a continous U(1) symmetry.  This qualitative
picture leaves several issues unresolved that numerical simulations have
not been able  to resolve  either: 
\begin{itemize}
\item What is the nature of the two phase transitions for $p \gtrsim 5$?

\item What is the nature of the relevant topological excitations
in each phase?

\item What is the physical origin of the critical (massless) phase?

\item What is the minimum $p$ after which the transitions are of the BKT
type? 

\end{itemize}

To understand qualitatively  the nature of the phases of the model, 
consider the ground state of the  self-dual quantum Hamiltonian \(H_p\)
defined in Eq. \eqref{Hpself-dual},  in the large (low-temperature) and
small  (high-temperature) $K_1$ limits ($a_0(K_2)=0$).  Let us start
with the  broken $\Z_p$ symmetry, low temperature sector that
corresponds to the line $(K_1,a_1(K_2)=0)$.   Then,  the $p$-fold
degenerate subspace of ground states is trivial to describe  in terms of
the simultaneous  eigenvectors of the $U_i$ of Eq. \eqref{eigenvofU},
\begin{eqnarray}
\ket{\Psi_0^s}= \prod_{i=1}^N \ket{s_i}, \ \mbox{ with same }  s  \mbox{
for all } i .
\end{eqnarray}
The ground state energy is $E_0=-K_1 N$ for periodic boundary
conditions  ($E_0=-K_1 (N-1)$  for open boundary conditions), and
$\langle \Psi_0^r|\Psi_0^{s}\rangle=\delta_{rs}$ \cite{chargeconjgs}. 
The fully disordered, high-temperature phase is defined by the  sector
$(K_1=0,a_1(K_2))$. The ground state ($E_0=-2a_1(K_2) N$) is unique and 
given by
\begin{eqnarray}
\ket{\Phi_0}= \prod_{i=1}^N \ket{\tilde{0}_i} ,
\end{eqnarray}
(in terms of the eigenstates of $V_i$,  Eq. \eqref{eigenvofV}), and
satisfies $\mathcal{C}_0 \ket{{\Phi}_0}= + \ket{{\Phi}_0}$.  It is
difficult to obtain exact results for arbitrary couplings. There is,
however, an interesting exact relation that holds at the  self-dual line
\(K_1=2a_1(K_2)\equiv K^*\) and follows from the fact that  the
self-duality unitary \(\mathcal{U}_{\sf d}\) becomes a new symmetry of 
the problem on that line. It is clear from Eq. \eqref{Hpself-dual} that 
$H_p[K^*]=-K^*(H_U+H_V)$, and  \(\mathcal{U}_{\sf d}H_U \mathcal{U}_{\sf
d}^\dagger=H_V\), \(\mathcal{U}_{\sf d}H_V \mathcal{U}_{\sf
d}^\dagger=H_U\) \cite{dualities}. Since  \([H_p[K^*],\mathcal{U}_{\sf
d}]=0\), we can choose the  energy eigenstates \(|\Psi_n\rangle,\
n=0,1,\cdots\), to be also eigenstates of \(\mathcal{U}_{\sf d}\),
\(\mathcal{U}_{\sf d}|\Psi_n\rangle =e^{\im\phi_n}|\Psi_n\rangle\). Then
\begin{equation}
E_n=\langle \Psi_n|H_p[K^*]|\Psi_n\rangle=
-2K^*\langle \Psi_n|H_U|\Psi_n\rangle=-2K^*\langle \Psi_n|H_V|\Psi_n\rangle.
\end{equation}

For $2\leq p \leq 4$, the $p$-clock model is {\it exactly} self-dual and
the  transition from the ferromagnetic to the disordered phase happens
at the  self-dual point $K_1=2 a_1(K_2)$.  For $p \geq 5$, Eq.
\eqref{exacthp} shows that  the $p$-clock model is no longer exactly
self-dual,  but the self-dual approximation of Eq. \eqref{Hpself-dual} 
or \eqref{t_sdpclock} allow us to establish the following {\it self-dual
equation} for arbitrary $p$
\begin{eqnarray}
\hspace*{-0.5cm} \frac{b_1}{b_0}=e^{K_2 (\cos \frac{2\pi}{p}-1)}=\half
\frac{\partial \ln B_p(a_1)}{\partial a_1}, \ \ \mbox{ where } B_p(a_1)
= \sum_{m=0}^{p-1}e^{2a_1\cos (\frac{2\pi}{p} m)}.
\label{SD}
\end{eqnarray}
{}From the  self-dual condition $K_1=2a_1(K_2)$ one can determine   the
{\it self-dual temperature} $T^*$. The self-dual point is a point of 
non-analyticity of the free energy for $2\leq p \leq 4$, but for  $p
\geq 5$ it is analytic.  Some values are indicated in  Table
\ref{tbl1.1}, assuming  isotropic couplings $K_1=K_2=J/(k_B T)$. It
follows from very general considerations  (see Section 8  of Ref.
\cite{dualities}) that  the two  critical points $c_1$ and $c_2$
bounding the self-dual point when $p \geq 5$  are exactly related by
\begin{eqnarray}
 \frac{K_1}{2a_1(K_2)} \Big \rfloor_{c_1}  \cdot  \frac{K_1}{2a_1(K_2)}
\Big \rfloor_{c_2}=1 .
\label{SDeqp5}
\end{eqnarray} 

\begin{table}[htb]
\begin{center}
\caption{Critical, $T_c$,  and self-dual, $T^*$, temperatures. For $p\geq
5$, there  are two critical temperatures. The lowest one, $T_{{\sf
BKT}}^{(1)}$,  goes to zero when $p\rightarrow \infty$, as $T_{{\sf
BKT}}^{(1)} \sim 1/p^2$, and the highest critical temperature $T_{{\sf
BKT}}^{(2)} \sim {\cal O}(1)$.}
\begin{tabular}{ccc}
\hline
$p$ & $T_c$ & $T^*$\\
& $[J/k_B]$ &  $[J/k_B]$\\ \hline 
2& $2/\ln (1+\sqrt{2})$& $2/\ln (1+\sqrt{2})$\\
3 & $3/(2\ln (1+\sqrt{3}))$&$3/(2\ln (1+\sqrt{3}))$ \\
4 & $1/\ln (1+\sqrt{2})$ & $1/\ln (1+\sqrt{2})$\\
6& $\cdots$ & $1/(2 \ln (2 \cos(\frac{\pi}{9})))$ \\
large $p$ & $\cdots$ & $2\pi/p$ \\ \hline
\end{tabular}
\label{tbl1.1}
\end{center}
\end{table}

It is interesting to analyze the large-$p$  limit of the self-dual
Eq.   \eqref{SD}. In that limit
\begin{eqnarray}
\lim_{p \rightarrow \infty} \frac{1}{p}B_p(a_1)=I_0 (2a_1) \ , \ 
\lim_{p \rightarrow \infty} \frac{1}{2p}\frac{\partial
B_p(a_1)}{\partial a_1}=I_1 (2a_1) ,
\end{eqnarray}
and from the asymptotic expansion of the  modified Bessel functions
$I_{0,1}$, Eq. \eqref{asympBess},  one gets the relation between the
transfer matrix  and direct couplings
\begin{eqnarray}
\frac{1}{2a_1(K_2)}= \frac{4\pi^2}{p^2} K_2 .
\label{largepdual}
\end{eqnarray}
One can then use Eq. \eqref{SDeqp5} to obtain a relation between the two
critical  temperatures $T_{{\sf BKT}}^{(1)}$ and $T_{{\sf BKT}}^{(2)}$
($J_1=J_2=J$) 
\begin{eqnarray}
k_B T_{{\sf BKT}}^{(1)}= \frac{4\pi^2}{p^2} \frac{J^2}{k_B T_{{\sf
BKT}}^{(2)}}.
\label{dual_t1t2}
\end{eqnarray}
The Peierls argument developed in  \ref{PeierlsApp},  on the other hand,
provides a rigorous scaling for the lowest transition  temperature,
$T_{{\sf BKT}}^{(1)}  = {\cal{O}}(1/p^{2})$, as $p \to \infty$.  Thus
this lowest critical temperature vanishes for the classical XY model
(where a broken-symmetry phase is not allowed) and,  according to Eq. 
(\ref{dual_t1t2}),  $T_{{\sf BKT}}^{(2)} \sim {\cal O}(1)$ for  large
$p$.  The self-dual temperature \(T^*\) has its own ``intermediate"
scaling with $p$,  $k_B T^*=(2\pi/p)  J$, so that it also vanishes in
the $p \rightarrow \infty$ limit. This is to be expected since the  XY
model is not self-dual, nor has a natural self-dual approximation.

\begin{figure}[thb]
\begin{center}
\includegraphics[angle=0,width=12cm]{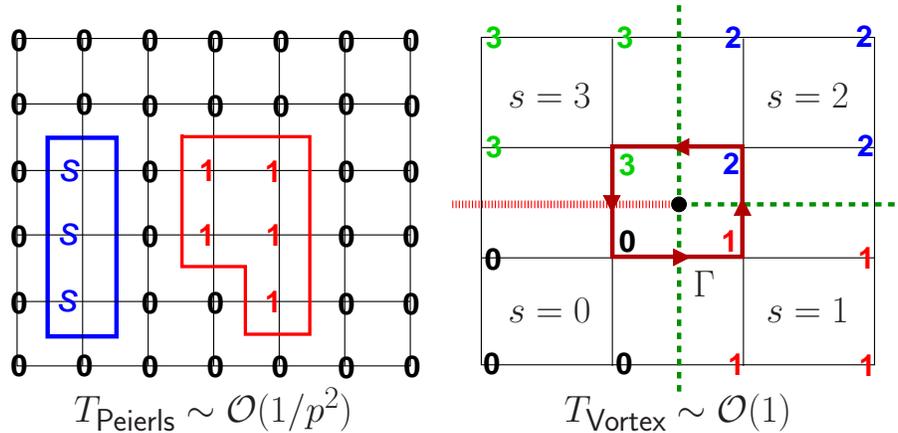}
\end{center}
\caption{Two types of topological excitations: domain wall (left panel)
and {\it discrete}  vortex-like  (right panel) excitations. Integer
numbers $s=0,1,2,3$ indicate the angle variables  $\theta=2\pi s/p$. For
$p\geq 5$ the energy cost of domain walls depends  on $|\Delta s|$ as
opposed to the $2 \leq p \leq 4$ case where the cost is independent  of
$|\Delta s|$.  Vortex-like configurations start appearing at $p=5$, and
the example above shows  a vortex in $\mathbb{Z}_5$ of strength $k=1$.}
\label{fig_pexc}
\end{figure}

To understand what makes $p\geq 5$ different from $p < 5$ and explain
the appearance of the intermediate critical phase, one needs to analyze
the nature of the topological  excitations. For $p\geq 5$  there are two
main types of topological excitations (see  Fig. \ref{fig_pexc}): (i)
domain wall excitations
%, characterized by configurations 
that dominate the low-temperature  physics, and (ii) discrete
vortex-like excitations of relevance in the critical and
high-temperature  phases.  The key distinction between the two  is that
domain walls exist for any \(p\), while vortex-like excitations  do not
exist for $2\leq p \leq 4$, becoming manifest only for  $p\geq5$. Also,
if $2 \leq p \leq 4$  the energy cost to create a domain wall is
independent  of $|\Delta s|=|s_{\sf in}-s_{\sf out}|$, with $s_{\sf
in(out)}$ indicating the angular configurations at the  two sides of the
wall. This changes for $p \geq 5$, allowing for twists of the spin of
size $|\Delta s| =2, \cdots, (p-2)$.

In the Peierls argument provided in \ref{PeierlsApp}, the upper
bound for the probability of having a domain wall corresponds to a
change of orientation between  two domains of $(\pm 2 \pi/p)$ (or
equivalently $|\Delta s| \sim {\cal O}(1)$).  Such a change {\em always
appears} in all domain walls in systems with $p=2,3,4$.  This, however,
is not the case for $p \geq 5$ where there exist general domain wall
topologies that do not allow for a uniform twist of the angle between
neighboring domains. In such instances, $|\Delta s|$ can be ${\cal
O}(p)$ and, as shown in Fig.  \ref{fig_pexc}, two types of excitations
are generally possible.  More precisely, in $p \ge 5$ systems, a {\em
vorticity} arises. The topological invariant  characterizing these
configurations, that we call {\it discrete winding number} $k$,  is
given by the circulation sum
\begin{eqnarray}
k = \frac{1}{2\pi}\sum_\Gamma \hspace*{-0.65cm}{\bigcirc} 
\hspace*{0.3cm} \Delta \theta_{\r \r'},
\label{mr}
\end{eqnarray}
taken around an oriented loop $\Gamma$, with  the argument $\Delta
\theta_{\r \r'} \in [-\pi, \pi) $ given by 
\begin{eqnarray}
\Delta \theta_{\r \r'} \equiv (\theta_{\r'} - \theta_{\r})  \mbox{
mod}(2 \pi).
\end{eqnarray}
In contrast, the sum in Eq. \eqref{mr} {\it should not be taken}
mod(\(2\pi\)) but just as an ordinary sum of real numbers (otherwise it
would vanish identically). To make this definition lucid, consider a
($p=5$) configuration such as the one  shown in Fig. \ref{fig_pexc}, 
with $\theta = 2 \pi s/p$, and a loop $\Gamma$. Herein, the circulation
sum explicitly reads $(\Delta \theta_{01} + \Delta \theta_{12}  + \Delta
\theta_{23} + \Delta \theta_{30})= 2\pi/5 + 2\pi/5 + 2\pi/5 + 4\pi/5 = 2
\pi$.  Thus, the configuration  in Fig.  \ref{fig_pexc} has a vortex of
strength $k=1$  at its origin. As $\theta_{0} - \theta_{3} = - 6 \pi/5$,
we set $\Delta \theta_{30} = 4\pi/5$. This shift in the value of 
$\Delta \theta_{\r \r'}$ (that must lie in the interval $[-\pi, \pi)$)
leads to the non-zero value of $k$ in this case. 

We may now use energy-versus-entropy balance considerations to argue for
the relevance of these  topological excitations in establishing the two
phase transitions.  The Peierls argument  presented in  \ref{PeierlsApp}
rigorously establishes that  domain walls oriented relative to one
another by the minimal energy cost  (i.e., twists of $(\pm 2 \pi/p)$)
are responsible for the existence of a low-temperature  ferromagnetic
broken $\mathbb{Z}_p$ symmetry phase.  Both the energy penalties and
entropic costs associated with such  {\it minimal cost} domain walls
scale with $\ell$ (the domain wall length), and  the analysis leads to a
transition temperature that behaves as $k_B T_{{\sf BKT}}^{(1)} \sim 
(1-\cos \frac{2\pi}{p}) J$.  The second transition temperature, $T_{{\sf
BKT}}^{(2)}$ ($p \geq 5$),  is  associated with the proliferation of
vortex-like excitations, and as indicated in  \ref{PeierlsApp} should
scale as  $T_{{\sf BKT}}^{(2)} \sim {\cal O}(1)$. Note that this
energy-versus-entropy balance argument does not  rely on the existence
or non-existence of the self-dual property of  the model. 

It is important to mention that while the physics of the low-temperature
phase is  associated to the exact discrete $\mathbb{Z}_p$ symmetry,  the
existence of vortex-like excitations is  directly related to the {\it
emergence} of the continuous symmetry  $\hat{\cal U}_\phi$ unveiled in 
Section \ref{symmetriesp}. This U(1) symmetry  becomes {\it more exact}
at high temperatures  ($T \gtrsim T^*$ or $2 a_1 \gtrsim K_1$), for a
fixed  $p \geq 5$, or it is exact at any temperature when $p \rightarrow
\infty$.  Thus, the physical origin of the critical phase and the {\it
extended  universality} concept introduced in  Ref. \cite{Lapilli} is
simply our emergent  $\hat{\cal U}_\phi$ continuous symmetry.

What is the nature of the phase transitions when $p\geq 5$?  Given the
current debates \cite{Borisenko}, it is important to say what we mean by
a ``BKT-type phase transition''. We simply mean a  transition
characterized by an essential singularity in the free energy (or the
ground state energy in a quantum model). This includes those cases where
there is an essential  singularity but, for instance,  the correlation
function exponent $\eta \neq 1/4$  (1/4 is the exponent for the XY model
\cite{NO}). It is very difficult to prove analytically the existence of
an essential singularity but numerical simulations seem to indicate that
for $p\geq5$ the two phase transitions are continuous with continuous
derivatives \cite{Borisenko}, supporting the BKT scenario. Moreover, our
new self-duality argument proves that the two transitions must be of the
same nature \cite{dualities}.  In other words, if there is an essential
singularity (the function and all its derivatives remain  continuous) in
the free energy at  $T_{{\sf BKT}}^{(1)}$, then,  there should be the
same type of singularity at  $T_{{\sf BKT}}^{(2)}$  \cite{dualities}.
This, of course, does not mean that the self-duality fixes the  value
of, for instance, $\eta$, to be the same at the two transition points.

In what follows, we will show that when $p\geq 5$, the temperature 
region  $T^{(1)}_{\sf BKT} \le T \le T^{(2)}_{\sf BKT}$ must be critical
(massless) with algebraic correlations.  The assumptions are:
(1) The phase transitions at $T^{(1)}_{\sf BKT}$ and  $T^{(2)}_{\sf
BKT}$ are continuous, and 
(2) for large separations $|\r -\r'|$, the two-point correlation 
function   $G(|\r -\r'|,T)=\langle \cos(\theta_\r -\theta_{\r'})
\rangle$ is of the canonical Ornstein-Zernike-like  form
\begin{eqnarray}
G(|\r-\r'|,T) \simeq A \,  \frac{e^{-|\r-\r'|/\xi}}{|\r-\r'|^{\eta}}+M^2,
\end{eqnarray}
with $M$ representing the order parameter, i.e., magnetization, $\xi$
the correlation length,  $\eta$ an anomalous exponent, and $A$ an
amplitude. In principle, all of these quantities  $(M, \xi, \eta,$ and
$A$) are functions of temperature $T$.  

We start by demonstrating that $G(|\r -\r'|,T)$ is a monotonically
decreasing function of  $T$ for any $p$. To this end, we derive in
\ref{appFF} a Griffiths'-type inequality  as in  general ferromagnetic
systems \cite{second}. Assume,  for simplicity, the uniform case
$K_\mu=K > 0$. Then, it is straightforward to  show (see \ref{appFF}) that 
\begin{eqnarray}
\hspace*{-1.5cm}
\partial_K G(|\r-\r'|,T) \ge 0,
\label{mono}
\end{eqnarray}
or, equivalently, $\partial_T G(|\r-\r'|,T) \leq 0$, which proves that
$G$ is a monotonically  decreasing function of temperature 
($0 \leq G(|\r-\r'|,T) \leq 1$). 

Now, if $G(|\r-\r'|,T)$ is monotonically decreasing with temperature $T$
for any fixed separation $|\r-\r'|$ then, in the absence of
magnetization (i.e., when $M=0$) the correlation length $\xi$ must be a
monotonically decreasing function of temperature. The proof of this
assertion is trivial. Consider two rather general temperatures in this
region $T_a, T_b$, such that  $T^{(1)}_{\sf BKT} \leq T_{a}<T_{b} \leq
T^{(2)}_{\sf BKT}$, then the ratio  of the corresponding asymptotic
correlation functions is 
\begin{eqnarray}
\frac{G(|\r-\r'|,T_{b})}{G(|\r-\r'|,T_{a})} = \frac{A_{b}}{A_{a}}
\frac{e^{|\r-\r'| (\xi_{a}^{-1} -
\xi_{b}^{-1})}}{|\r-\r'|^{\eta_b-\eta_a}} \leq 1,
\label{ratioab}
\end{eqnarray} 
because of the monotonicity property of $G$.  This is only possible if
$\xi_{b}<\xi_{a}$ for otherwise for large $|\r-\r'|$, the ratio of the
two correlation functions would diverge exponentially in $|\r-\r'|$. As
$T_{a}$ and $T_{b}$ were rather general temperatures, it follows that
$\xi(T)$ is a monotonically decreasing function of the temperature $T$
so long as the magnetization $M =0$ (as it is for $T>T^{(1)}_{\sf
BKT}$).  Therefore, a divergence of the correlation length at
$T^{(1)}_{\sf BKT}$ and $T^{(2)}_{\sf BKT}$ implies that {\it within the
entire interval $T^{(1)}_{\sf BKT} \le T \le T^{(2)}_{\sf BKT}$, the
correlator $G$ is algebraic in $|\r-\r'|$}. We thus proved that there
must exist a {\it power law}, critical phase, between $T^{(1)}_{\sf
BKT}$ and $T^{(2)}_{\sf BKT}$.

\vspace*{0.7cm}
\noindent
{\bf Acknowledgements}

This work is partially supported by the National Science Foundation under 
Grant No. 1066293 and the hospitality of the Aspen Center for Physics. Fruitful 
discussions with C. D. Batista are acknowledged. 

%The physical origin of the BKT transition is the emergence of the $\hat{\cal
%U}_\phi$  continuous symmetry.

\begin{appendix}
\section{Exponential of shift operators}
\label{appF} 

In this appendix we collect some useful formulas to compute exponentials
and logarithms of shift operators. Let us start with the operator
\(L_z\), the infinitesimal generator of  translations on the circle. We
have the general relation
\begin{equation}\label{expl}
e^{\int_0^{2\pi}d\theta\ a(\theta)e^{-\im \theta L_z}}=
\int_0^{2\pi}d\theta\ b(\theta)e^{-\im \theta L_z}.
\end{equation}
Our goal is to compute \(a\) as a function of \(b\), and {\it
vice versa}. The first step is to notice that the Fourier transform
operator
\begin{equation}\label{circle_fourier}
F=\sum_{n\in\mathbb{Z}}\int_0^{2\pi}d\theta\ \  \frac{e^{-\im \theta
n}}{\sqrt{2\pi}} \ |n\rangle\langle \theta| ,
\end{equation}
puts \(L_z\) (and the expression of Eq. \eqref{expl}) in diagonal form,
\begin{equation}\label{lz_diagonal}
Fe^{-\im \theta L_{z}}F^\dagger=\sum_{n\in\mathbb{Z}}|n\rangle\langle n|
e^{-\im \theta n}.
\end{equation}
This, together with the orthogonality relation 
\(2\pi\delta(\theta'-\theta)\!=\! \sum_{n\in \mathbb{Z}} e^{\im
(\theta'-\theta)n}\),  leads to
\begin{eqnarray}
b(\theta)&=&\frac{1}{2\pi}\sum_{n\in\mathbb{Z}}\ e^{\im \theta n} 
e^{\int_0^{2\pi}d\theta'  a(\theta')e^{-\im \theta' n}},\label{lab1}\\
a(\theta)&=&\frac{1}{2\pi} \sum_{n\in\mathbb{Z}}\ e^{\im \theta n}\ln 
\left(\int_0^{2\pi}d\theta' b(\theta')e^{-\im \theta' n} \right).
\label{lab2}
\end{eqnarray}

In Section \ref{tp} we introduced a diagonal matrix $U$, and a shift
operator \(V^\dagger\)  that describes translations in a $p$-points
discretization of the circle. This shift operator plays a role  similar
to that of \(L_z\) (actually, \(e^{-\im \theta L_z}\)). Now we have the
general relation
\begin{equation}
e^{\sum_{m=0}^{p-1}\ a_mV^m}=\sum_{m=0}^{p-1}\ b_mV^{\dagger m},
\end{equation}
that should be compared to Eq. \eqref{expl}. Our goal is to find
closed-form expressions for the coefficients \(a_m\) in terms of \(b_m\)
and {\it vice versa}.  The unitary transformation that diagonalizes
\(V^\dagger\) is now given by the {\it discrete} Fourier transform of
Eq. \eqref{wga_aut}. Putting  these pieces together, we get the solution
to our problem,
\begin{equation}
b_m=\frac{1}{p}\tr \left[U^{m}\ e^{\sum_{l=0}^{p-1}\ a_lU^{\dagger l}}
\right],\ \ \ \ \ \ a_m=\frac{1}{p}\tr
\left[U^{m}\ln\left(\sum_{l=0}^{p-1}\ b_l U^{\dagger l}\right) \right]\ .
\end{equation}
As seen by expanding the trace, these equations are closely related to  
\eqref{lab1} and \eqref{lab2}.

In physical applications, the \(a_m\) are Hermitian-symmetric,
\(a_{p-m}=a_m^*\) (to guarantee that \(\sum_{m=0}^{p-1}\ a_mV^m\)  is a
Hermitian operator), and the \(b_m\) are real and positive. Thus, it is
convenient to assume that both set of coefficients satisfy
\(a_{p-m}=a_m\), \(b_{p-m}=b_m\), and the relations between them
simplify to 
\begin{eqnarray}
b_m&=&\frac{1}{p}\sum_{s=0}^{p-1}\ \cos (\frac{2\pi ms}{p}) \
e^{\sum_{l=0}^{p-1}\  a_l\cos(\frac{2\pi l
s}{p})},\label{cos_bs_and_as}\\
a_m&=&\frac{1}{p}\sum_{s=0}^{p-1}\ \cos(\frac{2\pi m
s}{p})\ln\left(\sum_{l=0}^{p-1}\  b_l\cos(\frac{2\pi l s}{p})\right).
\end{eqnarray}
These are the expressions that are most useful in physical applications. 

Suppose next that \(b_m=e^{K\, u(m)}\), where \(K\) is a positive
constant, and \(u(m)\) is a real function of \(m=0,\cdots,p-1\) (for
example, \(u(m)=\cos(\frac{2\pi m}{p})\) for the classical $p$-clock
model). We would like to study the behaviour of the \(a_m\) to
next-to-leading order in \(K\), in the limit that \(K\) grows very large
(this could happen at low temperature). Notice that in this limit
\begin{equation}\label{approx_cs}
\sum_{l=0}^{p-1}\ b_l\cos(\frac{2\pi l s}{p})\approx  e^{K
u(0)}\left(1+2e^{K (u(1)-u(0))}\cos(\frac{2\pi s}{p})\right)
\end{equation}
to next-to-leading order, assuming that the inequalities
\begin{equation}
0>(u(1)-u(0))>(u(2)-u(0))>\cdots
\end{equation}
hold. The factor two in Eq.  \eqref{approx_cs} is due to  the symmetry
\(u(p-l)=u(l)\). Replacing expansion \eqref{approx_cs} into Eq. 
\eqref{cos_bs_and_as} leads to
\begin{equation}\label{plarge_K}
a_m\approx K
u(0)\delta_{m,0}+e^{K(u(1)-u(0))}(\delta_{m,1}+\delta_{m,p-1}), \ \ \ \
\ K\rightarrow\infty,
\end{equation}
where we have used  \(\ln(1+x)\approx x\), and the  orthogonality
relation
\begin{equation}\label{orthocos}
\frac{1}{p}\sum_{s=0}^{p-1}\ \cos(\frac{2\pi m s}{p})\cos(\frac{2\pi s
l}{p})= \frac{1}{2}\left(\delta_{m,l}+\delta_{m,p-l}\right).
\end{equation}

\section{Duality of the XY model to $q$-deformed bosons}
\label{xyqbosons}

In this section we study a duality that illustrates the essential
differences  between compact, $\hat{\theta}$, and non-compact, 
$\hat{x}$, degrees of
freedom. The algebraic tool of choice is the \(q\)-oscillator algebra
\cite{rideau}, specified by a positive real number \(q\), a creation
operator \(a^\dagger\), its Hermitian conjugate  \(a\), and a Hermitian
operator \(\hat{n}\), satisfying
\begin{eqnarray}\label{qalg}
[\hat{n},a]=-a,\ \ \ \ [\hat{n},a^\dagger]=a^\dagger,\ \ \ \ 
aa^\dagger-qa^\dagger a=q^{-\hat{n}}. 
\end{eqnarray}
If \(q=1\), this algebra reduces to the standard harmonic oscillator
algebra, that is isomorphic to the Heisenberg algebra of translations 
on the line, $[\hat{x},\hat{p}]=\im$, provided \(a=(\hat{x}+ \im
\hat{p})/\sqrt{2}\), \(a^\dagger =(\hat{x}- \im \hat{p})/\sqrt{2}\). It
was pointed out in Ref. \cite{kowalski} that  the mapping 
\begin{eqnarray}\label{circleto_defboson}
L_z &\mapsto& -\hat{n}+\ln\sqrt{2\sinh(1)},\label{ddefbosons}\\
e^{\im\hat{\theta}} &\mapsto& \sqrt{2\sinh(2)}\, a \, e^{-\hat{n}},\ \ \ \ 
e^{-\im \hat{\theta}}\mapsto\sqrt{2\sinh(2)} \,
e^{-\hat{n}}a^\dagger ,\nonumber
\end{eqnarray}
affords a representation of the algebra of translations in the circle
$[L_z,\ e^{\pm\im\hat{\theta}}]=\pm e^{\pm\im\hat{\theta}}$, {\it
provided that we set \(q=e^{-2}\) in} Eq. \eqref{qalg}.

Clearly, we can extend the mapping of Eq. \eqref{circleto_defboson} to a
duality isomorphism \(\Phi_\d\) for the XY model. The dual transfer
operators read
\begin{eqnarray}
T_2^D&=&\prod_{i=1}^{N}\  \sqrt{2\sinh(1)}\int_0^{2\pi}d\theta\
e^{K_2\cos\theta}e^{\im\theta \hat{n}_i},\\
T_1^D&=&\prod_{i=1}^{N-1} \
\exp\left[K_1\sinh(2)(a_{i+1}e^{-(\hat{n}_{i+1}+\hat{n}_i)}a_i^\dagger+{\sf
h.c})\right]. \nonumber
\end{eqnarray}
To compute \(\pf_{\sf XY}^D\), one should take the trace in the
eigenbasis of \(\hat{n}_i\). This basis is  described in Ref.
\cite{rideau}.

This description of the XY model in terms of \(q\)-deformed bosons with
\(q=e^{-2}\) suggests that the algebra of Eq. \eqref{qalg} affords a
continuous interpolation between the XY model and ordinary phonons 
(characterized by \(q=1\)), but this is not the case:
The XY model belongs to a representation of the algebra of Eq. \eqref{qalg}
that is inequivalent to that  describing phonons (i.e., canonical bosons). The
reason is that  Eq. \eqref{h_circle} is not enough to specify the
algebra of translations in the circle. We must also have that
\begin{equation}
e^{\pm\im\hat{\theta}} e^{\mp \im\hat{\theta}}=\mathds{1}.
\end{equation}
The mapping of Eq. \eqref{circleto_defboson} will respect this
constraint only if \(a, a^\dagger\) satisfy 
\begin{equation}
aa^\dagger-q^{-1}a^\dagger a=0
\end{equation}
at least for \(q=e^{-2}\),  {\it including the relations listed in} Eq.
\eqref{qalg}. But the resulting set of four relations becomes
inconsistent at \(q=1\). This shows that the \(q\)-oscillator algebra
cannot interpolate continuously between canonical bosons and compact
excitations.

\section{The Villain and its dual solid-on-solid models}
\label{villain}

The Villain model \cite{NO}
\begin{equation}
\pf_{\sf V}[K_\mu]=\sum_{\{n_{(\r,\mu)}\}}\sum_{\{\theta_\r\}}
\exp\left[\sum_\r\sum_{\mu=1,2}\frac{K_\mu}{2}(\theta_{\r+\bm{e_\mu}}-
\theta_\r-2\pi n_{(\r,\mu)})^2 \right],
\end{equation}
was introduced in Ref. \cite{villain} to provide a Gaussian
approximation to the XY model that preserves the essential property of
compacticity, and is a good approximation at sufficiently low
temperatures.  We now show, by using our bond-algebraic approach,  that
it is dual to the solid-on-solid (SoS) model of the roughening
transition,
\begin{equation}
\pf_{\sf SoS}[K_\mu]=\sum_{\{m_\r\}}\ \exp\left[\sum_\r\sum_{\mu=1,2}\ 
K_\mu^{-1} (m_{\r+\bm{e_\mu}} -m_\r)^2\right],
\end{equation}
characterized by integer-valued degrees of freedom \(m_\r\in\mathbb{Z}\)
\cite{NO}. We work directly in the thermodynamic limit, 
\(N\rightarrow \infty\), to avoid dealing with boundary terms.

The transfer operator \(T_{\sf SoS}=T_2T_1\) for the SoS model
can we written  as 
\begin{equation}
T_1=\prod_i\ e^{\frac{K_1}{2}(X_{i+1}-X_i)^2},\ \ \ \
T_2=\prod_i\ \sum_{m}\ e^{\frac{K_2}{2}m^2} R^{\dagger m}_i,
\end{equation}
in terms of the operators \(X_i,\ R_i,\ R_i^\dagger\)
defined in Eq. \eqref{integer_circle}. Now, however,
\(i\in \mathbb{Z}\) labels the sites of an infinite straight line.  
The duality of bond algebras 
\begin{equation}
X_{i}-X_{i-1}\dual L_{z,i},\ \ \ \ R_i\dual e^{\im(\hat{\theta}_{i+1}-
\hat{\theta}_i)},\ \ \ \ R_i^\dagger\dual e^{-\im(\hat{\theta}_{i+1}-
\hat{\theta}_i)}
\end{equation} 
affords a dual representation of \(T_{\sf SoS}\),
\begin{equation}
T_1^D=\prod_i\ e^{\frac{K_1}{2}L_{z,i}^2},\ \ \ \ T_2^D=\prod_i\
\sum_{m}\ e^{\frac{K_2}{2}m^2} e^{-\im (\hat{\theta}_{i+1}-
\hat{\theta}_i) m} ,
\end{equation}
in terms of compact degrees of freedom. The next step is to compute
\(\pf_{\sf SoS}^D=\tr [(T_2^D T_1^D)^N]\) in the basis introduced in Eq.
\eqref{circle_kets}.

\(T_2^D\) is already diagonal in that basis
\begin{equation}
T_2^D|\theta\rangle= \prod_i\ \sum_{m}\ e^{\frac{K_2}{2}m^2}  e^{-\im
(\theta_{i+1,j}-\theta_{i,j}) m} \ |\theta\rangle .
\end{equation}
At this point we could proceed by analogy to previous sections and 
rewrite this expression in terms of an interaction potential
\(V_K(\theta)\equiv -\ln \sum_{m}\ e^{\frac{K_2}{2}m^2} e^{-\im \theta
m}\), but this will not turn out be the most convenient approach.
Instead, let us proceed to compute the matrix elements of \(T_1^D\).
This task reduces to computing the matrix elements of a one-body
operator,
\begin{equation}
\langle\theta_i'|^{\frac{K_1}{2}L_{z,i}^2}|\theta_i\rangle = 
\frac{1}{2\pi}\sum_{m_i}\ e^{\frac{K_1}{2}m_i^2}e^{-\im
(\theta_i'-\theta_i) m_i},
\end{equation}
which results from recalling that the orthonormal states of  \(L_{z,i}\)
are the plane waves \(\langle\theta_i|n_i\rangle=e^{\im \theta_i
n_i}/\sqrt{2\pi}\). Notice  that the function \(e^{\frac{K}{2}x^2}\) is
the Fourier  transform of \(e^{\frac{x^2}{2K}}/\sqrt{K}\). It then
follows that we can use Poisson's summation formula to write
\begin{equation}
\sum_{m_i}\ e^{\frac{K}{2}m_i^2}e^{-\im \theta m_i}=
\sqrt{\frac{2\pi}{K}}\sum_{m_i} e^{\frac{(\theta-2\pi m_i)^2}{2K}}. 
\end{equation}
 
Putting all the pieces together, we obtain
\begin{eqnarray}
\pf_{\sf SoS}^D&=&\left(\frac{2\pi}{K_2}\right)^N
\sum_{\{\theta_\r\}}\prod_\r\prod_{\mu=1,2}\ 
\sum_m\exp\left[\frac{K_\mu}{2} (\theta_{\r+\bm{e_\mu}}-\theta_\r-2\pi
m)^2\right]\\
&=&\left(\frac{2\pi}{K_2}\right)^N\sum_{\{n_{(\r,\mu)}\}}\sum_{\{\theta_\r\}}
\exp\left[\sum_\r\sum_{\mu=1,2}\frac{K_\mu}{2}(\theta_{\r+\bm{e_\mu}}-
\theta_\r-2\pi n_{(\r,\mu)})^2 \right]. \nonumber
\end{eqnarray}
The last expression is exactly \(({2\pi}/{K_2})^N\pf_{V}[K_\mu]\), and
thus the Villain model is dual to the SoS model. Notice the
reciprocal  relation between the couplings: The Villain
model is strongly coupled only if its dual SoS representation
is weakly coupled.

\section{The \(p\)-clock model for \(p=2, 3\), and $4$}
\label{p234}

Let us start with the simplest $p=2$ case. 
Then, $U_i=U_i^\dagger=\sigma^z_i$ and \(V_i=V_i^\dagger=\sigma^x_i\), 
and the transfer matrix \(T_p=T_2T_1\) of Eq. \eqref{t2tvop} reduces to
\begin{equation}
T_1=\prod_{i=1}^{N-1}\ e^{K_1 \sigma^z_i\sigma^z_{i+1}}, \ \ \ \ \ \  
T_2=\prod_{i=1}^{N}\left(e^{K_2}+e^{-K_2}\sigma^x_i\right). 
\end{equation}
This finite Ising
model is self-dual {\it up to boundary corrections}. The substitution
\(T_1\rightarrow e^{K_1\sigma^z_N}T_1\) renders the model {\it exactly}
self-dual for any $N$ \cite{dualities}. 

If \(p=3\), then \(V_i^{\dagger 2}=V_i\), and $T_2$ becomes 
\begin{equation}
T_2=\prod_{i=1}^{N}\left(e^{K_2}+e^{-\frac{K_2}{2}}(V_i+V_i^\dagger)\right).
\end{equation}
\(T_1\) is just as in Eq. \eqref{t2tvop}, with \(U\)s appropriate for
\(p=3\). It follows that if we introduce the boundary correction
\(T_1\rightarrow e^{\frac{K_1}{2}(U_N+U_N^\dagger)}T_1\)
\cite{dualities}, then \(T_2\dual T_1\) and \(T_1\dual T_2\), rendering 
\({\pf}_p\) exactly self-dual for any $N$. 

The case \(p=4\) is  special because it can
be mapped onto two decoupled Ising models \cite{class}. Since
\(V_i^2=V_i^{\dagger 2}\) in this case, \(T_2\) reads
\begin{equation}
T_2=\prod_{i=1}^N(e^{K_2}+V_i+V_i^\dagger+e^{-K_2}V_i^2).
\end{equation}
Moreover, it is easy to check that  the operator $2{\bf
C}_i=(\one+\sigma^z_{1,i})+(\one-\sigma^z_{1,i})\sigma^x_{2,i}$ (known
as a controlled-NOT gate in quantum computation) maps
\begin{eqnarray}
U_i&=&e^{\im \frac{\pi}{4}}{\bf C}_i \Big ( \frac{\sigma^z_{1,i}- \im 
\sigma^z_{2,i}}{\sqrt{2}} \Big ) {\bf C}_i,\\
{V}_i+{V}_i^\dagger&=&{\bf C}_i(\sigma^x_{1,i}+ \sigma^x_{2,i}) {\bf C}_i, 
\ \ \ \ \ \ V_i^2={\bf C}_i\sigma^x_{1,i}\sigma^x_{2,i}{\bf C}_i,\nonumber
\end{eqnarray}
and thus it follows that ${\bf C}=\prod_{i=1}^N {\bf C}_i$ maps 
\begin{eqnarray}
{\bf C}T_1{\bf C} &=& \prod_{i=1}^{N-1} e^{\frac{K_1}{2} (\sigma_{1,i}^{z} 
\sigma_{1,i+1}^{z}+  \sigma_{2,i}^{z}  \sigma_{2,i+1}^{z})} , \\ 
{\bf C} T_2 {\bf C}&=&\prod_{i=1}^{N}(e^{\frac{K_2}{2}}+e^{-\frac{K_2}{2}}\sigma^x_{1,i}) 
(e^{\frac{K_2}{2}}+e^{-\frac{K_2}{2}}\sigma^x_{2,i}),
\end{eqnarray}
that clearly defines two decoupled Ising models, with couplings 
that are half of those of the \(p=2\) model. In particular,
the \(p=4\) clock model is exactly self-dual provided \(T_1
\rightarrow e^{\frac{K_1}{2}(\sigma^z_{1,N}+\sigma^z_{2,N})}\ T_1\).

\section{Peierls argument for the $p$-plock model}
\label{PeierlsApp}

 We now use the Peierls argument to prove that there should be a broken 
symmetry phase (low-temperature ordered phase) in the $p$-clock  model
on the square lattice. The proof establishes the existence of a phase
transition at a temperature $T^{(1)}$ below which global $\mathbb{Z}_p$
symmetry is broken. For large $p \gg1$, $T^{(1)} = {\cal{O}}(1/p^{2})$. 

Specifically,  our objective is to show that if uniform boundary
conditions pertaining to one of the clock states $\theta = 2 \pi s/p$,
with $0 \le s\leq p-1$  a fixed integer, are applied on the boundary of the
square lattice, then there  {\em provably} exists a temperature $T_{\sf
Peierls}>0$ such that for temperatures $T<T_{\sf Peierls}$, spontaneous
symmetry breaking (SSB) of the global $\mathbb{Z}_p$ symmetry arises. 
(In the context of our discussions thus far, $T_{\sf Peierls} <
T^{(1)}$; asymptotically, for large $p$, both temperatures scale as
$1/p^{2}$.) By SSB in this context, we refer to the lifting of the
symmetry triggered by applying the uniform boundary conditions at
spatial  infinity. That is, when the aforementioned boundary conditions
are introduced then, for $T<T_{\sf Peierls}$, the probability
distribution ${\cal{P}}(\theta_{{\bf{0}}})$ for the angular orientation
of the spin at the origin ${\bf{S}}_{\bf{0}}$ is not symmetric between
the $p$ possible  values of $\theta_{{\bf{0}}}$. In particular, we will
demonstrate that ${\cal{P}}(\theta_{{\bf{0}}})$  is maximal when
$\theta_{{\bf{0}}}$ has an orientation that matches that on the
boundary, $\theta_{\bf{\infty}}$. In other words, for temperatures
$T<T_{\sf Peierls}$,
\begin{eqnarray}
{\cal{P}}(\theta_{\bf{0}} = \theta_{\bf{\infty}}) \ge \frac{1}{p}.
\label{SSB}
\end{eqnarray}
To prove this inequality, we note that 
\begin{eqnarray}
 {\cal{P}}(\theta_{\bf{0}} \neq \theta_{\bf{\infty}})  \le 
{\mbox{Prob(outer domain wall $\Gamma$)}},
\label{domain}
\end{eqnarray} 
where ${\mbox{Prob(X)}}$ denotes the probability of the set of events X.
The {\it domain wall} is defined as the boundary between differently
oriented spins. The logic underlying Eq. (\ref{domain}) is clear: if
$\theta_{\bf{0}} \neq \theta_{\bf{\infty}}$ then, by its very
definition, at least one domain wall must separate the spin at the
origin from  the spins on the boundaries of the lattice.

We now will bound the probability of having a particular domain  wall.
Specifically, let us denote by $\{C_{\alpha}\}$ the set of
configurations that have $\Gamma$ as the outer-most domain wall
surrounding the origin. That is, $\Gamma$ separates spins with an
orientation $\theta = \theta_{\infty}$ from those having another
(uniform) orientation $\theta_{\sf in}$. [Note that, generally, more
than one domain wall may be present and thus $\theta_{\sf in}$ need not
be the same as $\theta_{\bf{0}}$.] The upper bound on the probabilities
in Eq. (\ref{domain}) is a sum over the probabilities of having such
different outer-most domain walls $\Gamma$.  We furthermore define the
partition function
\begin{eqnarray} 
\pf_{\Gamma} = \sum_{\{C_\alpha\}} \exp[-\beta E_{\alpha}], 
\label{pfg}
\end{eqnarray}
where $E_{\alpha}  \equiv E(C_{\alpha})$ is the energy of the spin
configuration $C_{\alpha}$.  Thus, $\pf_{\Gamma}$ is smaller than the
total partition function $\pf_p$ of the system. This is so as
$\pf_{\Gamma}$ contains only a subset of the Boltzmann weights appearing
in $\pf_p$.  That is, in Eq. (\ref{pfg}) we sum  only over spin
configurations with at least one domain wall surrounding the origin. 

We next define a new configuration $\bar{C}_{\alpha}$ formed by rotating
all of the spins inside $\Gamma$ by a uniform angle $\Delta \theta$ such
that the outermost domain wall that surrounds the origin is removed. 
That is, for all spins ${\bf{S}}_\r$ that {(i)} lie inside the
region bounded by the domain wall $\Gamma$, we perform the
transformation  $\theta_\r \to (\theta_\r + \Delta \theta)$ with an
angle of rotation  
\begin{eqnarray}
\Delta \theta = \theta_{\infty}- \theta_{\sf in} \equiv \frac{2 \pi}{p}
\Delta s ,
\end{eqnarray}
where $\Delta s$ is an integer. {{(ii)}} All spins lying outside the
domain wall  $\Gamma$ have an orientation $\theta= \theta_{\infty}$;
these spins are not rotated. When present, any other more internal
domain walls will remain unchanged by this uniform rotation of all the
spins inside $\Gamma$.  In order to bound, from above, the probability
of having an outermost domain wall  $\Gamma$, we now consider
($E_{\bar{\alpha}}= E(\bar{C}_{\alpha})$)
\begin{eqnarray}
\pf_{\bar{\Gamma}} = \sum_{\{\bar{C}_{\alpha}\}} \exp[-\beta
E_{\bar{\alpha}}] .
\end{eqnarray}

The probability of having the domain wall $\Gamma$ is fixed by the 
ratio of the sum of Boltzmann weights associated with having the domain
wall $\Gamma$ divided by the sum of Boltzmann weights associated with
all spin configurations (i.e., the partition function $\pf_p$). As
$\pf_{\bar{\Gamma}}$ contains a sum only over a subset of all Boltzmann
weights that appear in $\pf_p$, we have
\begin{eqnarray}
{\mbox{Prob(outer domain wall $\Gamma$)}} = \frac{\pf_{\Gamma}}{\pf_p} 
\le \frac{\pf_{\Gamma}}{\pf_{\bar{\Gamma}}} = \frac{e^{-\beta E_{C_{1}}}
+  e^{-\beta E_{C_{2}}} + \cdots}{e^{-\beta E_{\bar{C}_{1}}} + e^{-\beta
E_{\bar{C}_{2}}} + \cdots}.
\label{bound}
\end{eqnarray}
The smallest energy difference between a configuration $C_{\alpha}$ and
$\bar{C}_{\alpha}$ is bounded by
\begin{eqnarray}
E_{C_{\alpha}} - E_{\bar{C}_{\alpha}} \ge  \ell (1- \cos \frac{2
\pi}{p}),
\label{eber}
\end{eqnarray}
where $\ell$ is the length of the domain wall $\Gamma$ (exchange
constants are set to unity,  $J_{\mu}=1$, $\mu=1,2$). As Eq. (\ref{eber}) applies
to all configuration pairs $C_\alpha$ and  $\bar{C}_\alpha$ that appear
in Eq. (\ref{bound}),
\begin{eqnarray}
\frac{e^{-\beta E_{C_{\alpha}}}}{e^{-\beta E_{\bar{C}_{\alpha}}}}  = 
e^{-\beta \ell (1- \cos \frac{2 \pi \Delta s}{p})} \le e^{-\beta \ell
(1- \cos \frac{2 \pi}{p})},
\end{eqnarray}
we have that
\begin{eqnarray}
\frac{\pf_{\Gamma}}{\pf_{\bar{\Gamma}}}  \le  e^{-\beta \ell (1- \cos
\frac{2 \pi}{p})}.
\label{ZZbound}
\end{eqnarray}
It is important to emphasize that when the bound of  Eq. (\ref{ZZbound})
is saturated, $|\Delta s| =1$. [Physically, for $p \gg1$, only such
domain walls (as opposed to far more energetically prohibitive domain
walls with $|\Delta s | = {\cal{O}}(p)$) may appear at sufficiently low
temperatures ($T \lesssim {\cal{O}}(1/p^{2})$).]

Returning to the probability that (at least) one domain wall surrounds
the origin in Eq. (\ref{domain}), we have that
\begin{eqnarray}
\label{dw}
{\mbox{Prob(outer domain wall $\Gamma$)}} \le \sum_{\ell} N_{\ell}
D_{\ell} ,
\end{eqnarray}
with $N_{\ell}$ denoting an upper bound on the number of domain walls of
perimeter $\ell$ that enclose the origin and $D_{\ell}$ an upper bound
on the probability of having a domain wall of length $\ell$. Inserting
Eq. (\ref{ZZbound}) while taking note of an upper bound  of $4 \times
3^{\ell-1}$ on the number of  non-backtracking walks of length $\ell$ on
the square lattice, and an upper bound of $(\ell/4)^{2}$ on the maximum
number of initial starting points for a walk of length $\ell$ that
surrounds the origin, we have
\begin{eqnarray}
{\mbox{Prob(outer domain wall $\Gamma$)}} \le && 
\hspace*{-0.4cm} \sum_{\ell \geq 4} \Big[
(\ell/4)^{2} \times 4 \times 3^{\ell-1}  e^{-\beta \ell(1- \cos \frac{2
\pi}{p})} \Big]  \nonumber \\ 
&& \equiv w(\beta, p) 
\end{eqnarray}
(the minimal domain wall on the square lattice has  length $\ell=4$).
The function $w$ is trivially bounded by performing the summation over
all natural numbers $\ell$
\begin{eqnarray}
w(\beta , p) \le  \sum_{\ell=1}^\infty \frac{3^{\ell}\ell^{2}}{12} 
e^{-\beta \ell(1- \cos \frac{2 \pi}{p})} = \frac{x(1+x)}{12(x-1)^{3}}
\equiv {\bar{w}}(\beta,p),
\label{wb}
\end{eqnarray}
with  $x=e^{\beta(1- \cos \frac{2 \pi}{p})}/3$.

In performing the summation in Eq. (\ref{wb}), we assumed a sufficiently
low temperature so that $x>1$, and  $\bar{w}$ is a monotonically decreasing
function of $\beta$. Notably, $\bar{w}$ can be made arbitrarily close to zero
for large enough $\beta$. Let us denote by $\beta_{\sf Peierls}$ the
solution to the equation $\bar{w}(\beta_{\sf Peierls},p)= \frac{p-1}{p}$.
Then, for $\beta > \beta_{\sf Peierls}$, the probability of the spin at
the origin being the same as that on the boundary is
${\cal{P}}(\theta_{\bf{0}} = \theta_{\infty}) >1/p$. In other words, for
$T < T_{\sf Peierls}$, we clearly have SSB. It is important to emphasize
that $T_{\sf Peierls}$ is only a lower bound to the transition
temperature, and the actual SSB occurs for $T^{(1)} >  T_{\sf Peierls}$. An
estimate for  $T_{\sf Peierls}$ resulting from this analysis is $T_{\sf
Peierls} \approx  (1-\cos \frac{2\pi}{p})/\ln 6$ ($\sim {\cal O}(1/p^2)$
for large $p$).  As in the lower bound derived herein,  the energy cost
for a domain wall is anticipated to determine the actual ordering
temperature. This bound  is {\em rigorous}. Physically, in Eq.
(\ref{dw}), the logarithms of the two terms $N_{\ell}$ and $D_{\ell}$
capture, respectively, bounds on the entropy and energy costs associated
with domain walls of length $\ell$. 

Discrete vortices such as the one shown in Fig. \ref{fig_pexc} with a
typical change of angle $\Delta \theta = {\cal{O}}(1)$ (or $|\Delta s| =
{\cal{O}}(p)$)  across the intersecting domain walls  that extend over a
linear distance  $\ell$ may entail, for all $p \geq 5$, an energy cost
that scales as $\ell$.  This is to be contrasted with the minimal energy
penalty associated with a difference in angle of $|\Delta \theta | = 2
\pi/p$ for which the corresponding energy penalty 
 as 
$\ell/p^{2}$ (and that physically sets the bounds that we derived in the
Peierls argument above). Thus, from energy-versus-entropy balance
considerations, the temperature below  which it is unfavorable to have
vortices is $T_{\sf Vortex} \sim {\cal O}(1)$ (or of order $J$): the
energy for such domain walls scales as $\ell$ as  does the entropy
associated with a network of  possible intersecting domain walls that
have a total length $\ell$. 

\section{Proof of monotonicity of the correlation function $G$}
\label{appFF}

We now prove Eq. (\ref{mono}) for large (yet finite) lattices.  In
finite size systems (no matter how large), there are no thermodynamic 
phase transitions. Thus, the free energy and all its derivatives
(including, in particular,  the two-point correlation function 
$G(|\r-\r'|,T)$) are analytic for all values of $\beta$.  We will prove
monotonicity by expanding (the analytic) $G$ as a power series in $\beta$,  which
is everywhere convergent, and illustrate that the coefficients multiplying each power of
$\beta$ are non-negative.  The uniformity of the sign of all contributions to the
series coefficients follows from repeated
applications of the identity 
\begin{eqnarray}
S_p(n)=\sum_{s=0}^{p-1} e^{{\sf i} \frac{2 \pi  n}{p} s} = p \,
\delta_{n,0} ,
\label{geometricseries}
\end{eqnarray}
with the Kronecker delta above defined mod($p$). That is,  $\delta_{0,0}
= \delta_{\pm p, 0} = \delta_{\pm 2 p,0} = \cdots=1$, otherwise it 
vanishes. 

Longhand, the correlator $G(|\r-\r'|,T)$ is given by ($K=\beta J \geq 0$)
\begin{eqnarray}
\langle \cos(\theta_\r- \theta_{\r'}) \rangle = \frac{\displaystyle
\sum_{\{\theta_\x\}} \cos(\theta_\r-\theta_{\r'}) \exp\left[\sum_{\x}
\sum_{\mu=1,2} K \, \cos(\theta_{\x+\bm{e_\mu}}-\theta_\x) \right ]}
{\pf_p[K]} ,
\label{ratioclassicalXYtheta}
\end{eqnarray}
with $\theta_\r=2\pi s_\r/p$.

We Taylor expand the argument of the exponential, i.e, $\exp[\beta A] =
\mathds{1}  + \beta A + (\beta A)^{2}/2! + \cdots$, in both the
numerator and the denominator of  Eq. (\ref{ratioclassicalXYtheta}), and
may represent pictorially the expansion terms by Feynman-type diagrams.
In this scheme, each appearance of $ \cos(\theta_{\x'}-\theta_\x)$
relates to an internal propagator linking sites $\x$ and  $\x'$. Similar
to perturbative schemes in the continuum,   where the number of
Feynman-type ``bubble diagrams'' in $\pf_p[K]$ with  nearest-neighbor
links that share common vertices with any given ``connected'' diagram 
which contains both the sites $\r$ and $\r'$ is negligible,
there is a near cancellation, in Eq. (\ref{ratioclassicalXYtheta}),  of  all
``bubble diagrams''. What remains from
the  ratio of Eq. (\ref{ratioclassicalXYtheta}) is the sum of all
diagrams in which sites $\r$ and $\r'$ are linked to each other by the
line stemming from $\cos(\theta_\r-\theta_{\r'})$ as well as internal
lines which  pass through  the points $\x =1,2,\cdots, m$.  Up to
(inherently positive) symmetry factors,  the numerical value of any such resulting
diagram is given by a sum of the form 
\begin{eqnarray}
\sum_{s_{\r}, s_1,s_2,\cdots,s_m, s_{\r'}=0}^{p-1}  \beta^{t_{\r 1} +
t_{\r 2} + \cdots + t_{12} + \cdots+  t_{m \r'}} 
\cos(\theta_\r-\theta_{\r'}) (\cos(\theta_\r- \theta_1))^{t_{\r 1}}
\nonumber \\ 
\times (\cos(\theta_\r- \theta_2))^{t_{\r 2}} \cdots (\cos(\theta_{1} -
\theta_{2}))^{t_{12}} \cdots (\cos(\theta_{m} -
\theta_{\r'}))^{t_{m\r'}} ,
\label{general_diagram}
\end{eqnarray}
where the integers $t_{ab}> 0$ represent the number of lines linking
sites $a$ and $b$.  It is simple to show that sums of the form of Eq.
(\ref{general_diagram}) are manifestly non-negative.  Replacing
$\cos(\theta_{a} - \theta_{b})$ by  $(\exp({\sf i}(
\theta_{a}-\theta_{b}))+ \exp({\sf i}( \theta_{b}-\theta_{a})))/2$,  a
sum of exponentials with positive weights,  Eq. (\ref{general_diagram})
reduces to a sum of individual products with positive weights, each
being of the form 
\begin{eqnarray}
\prod_{\x=\r,1,2,\cdots,m,\r'} S_p(n_\x) \geq 0,
\label{longproduct}
\end{eqnarray}
with $n_{a}$ set by sums of the powers $t_{ab}$ in Eq.
(\ref{general_diagram}).  Then,  it follows that when the ratio of Eq. 
(\ref{ratioclassicalXYtheta}) is expanded in $\beta$, i.e., 
$G(|\r-\r'|,T) = \sum_{t} a_{t} \beta^{t}$, the prefactor $a_{t}$
multiplying each individual  power $\beta^{t}$ is non-negative. As such,
$G(|\r-\r'|,T)$ is manifestly monotonic in $\beta$, i.e., 
$\partial_\beta G(|\r-\r'|,T) \geq 0$.

\end{appendix}


\begin{thebibliography}{00}
\bibitem{class}
For $p=4$  we may  represent the four equidistant spin orientations  on
the unit circle in terms of the normalized two component spin vectors
$(\chi^{1}_\r, \chi^{2}_\r)/\sqrt{2}$ at sites $\r$ with $\chi^{1,2}_\r=
\pm 1$. With this representation,  the isotropic partition function
reads
\begin{eqnarray}
\pf_{p=4}[K]=\sum_{\{\chi^{1}_\r, \chi^{2}_\r\}}\
\exp\left[\sum_{\r}\sum_{\mu=1,2} K (\chi_\r^{1} \chi^{1}_{\r +
\bm{e_\mu}} + \chi_\r^{2} \chi^{2}_{\r+ \bm{e_\mu}})/2 \right] ,
\end{eqnarray}
which shows that the $p=4$ clock model is equivalent to two decoupled
Ising systems.
\bibitem{chargeconjgs}
For $p\geq 3$,  ground states that are also eigenstates of $\mathcal{C}_0$, 
$\mathcal{C}_0 \ket{\tilde{\Psi}^s_\pm}=\pm\ket{\tilde{\Psi}^s_\pm}$, can
be  constructed:

(i) $p \in$ odd: 
\begin{eqnarray}
\begin{cases}
\ket{\tilde{\Psi}^0_+}=\ket{{\Psi}^0_0} \\
\ket{\tilde{\Psi}^s_\pm}=(\ket{{\Psi}^s_0} \pm
\ket{{\Psi}^{p-s}_0})/\sqrt{2} \ , \ \mbox{for } 1 \leq s \leq (p-1)/2 ,
\end{cases} \nonumber
\end{eqnarray}

(ii) $p \in$ even: 
\begin{eqnarray}
\begin{cases}
\ket{\tilde{\Psi}^0_+}=\ket{{\Psi}^0_0} \ , \
\ket{\tilde{\Psi}^{p/2}_+}=\ket{{\Psi}^{p/2}_0} \\
\ket{\tilde{\Psi}^s_\pm}=(\ket{{\Psi}^s_0} \pm
\ket{{\Psi}^{p-s}_0})/\sqrt{2} \ , \ \mbox{for } 1 \leq s \leq p/2-1
\end{cases} . \nonumber
\end{eqnarray}
\end{thebibliography}
\end{document}